\documentstyle[art8,aaspp4,flushrt,tighten,psfig]{article}

\slugcomment{Astrophysical Journal, 536, June 10, 2000 }

\begin{document}

\newcommand{\msun}{{\rm M_\odot}}
\newcommand{\kms}{{\rm{km}\,\rm{s}^{-1}}}

%
\def\ltsim{\raisebox{-.5ex}{$\;\stackrel{<}{\sim}\;$}}
\def\gtsim{\raisebox{-.5ex}{$\;\stackrel{>}{\sim}\;$}}
\def\hii{H\,{\sc~ii}}
\def\n{\footnotemark}
\def\o{\o}
%

\title{DYNAMICS OF WARM ABSORBING GAS\\ IN SEYFERT GALAXIES: NGC~5548}

\author{Mark C. Bottorff}
\affil{Department of Physics \& Astronomy, University of Kentucky,
Lexington, KY 40506-0055}

\author{Kirk T.\ Korista}
\affil{Department of Physics, Western Michigan University,
Kalamazoo, MI 49008}

\centerline{and}

\author{Isaac Shlosman}
\affil{Department of Physics \& Astronomy, University of Kentucky,
Lexington, KY 40506-0055}

\begin{abstract}

A hydromagnetic (MHD) wind from a clumpy molecular accretion disk
is invoked to explain observations of warm absorbing
gas at UV and X-ray energies in Seyfert~1 galaxies.  This paper focuses
on two important issues: (1) compatibility of kinematics and dynamics
of an MHD wind with the observed properties of warm absorbers; and (2)
the relationship between the UV and X-ray absorbing gases. We provide
an in-depth comparison between the MHD wind model and the well-studied
Seyfert~1 galaxy NGC~5548, which at high spectral resolution exhibits a
number of discrete UV absorption components. Model parameters used in
this study have been fixed by our previous work on the variability of
broad-line region in this object. The detailed UVX absorption structure
in NGC~5548 is modeled in order to infer the position, size, density
and temperature of each component, as well as the overall dynamics of
the absorbing gas.

We find that for NGC~5548: (1) the total column densities of
O{\sc~vii}, O{\sc~viii} and H, inferred from X-ray observations are
reproduced by constraining the UV ion column densities of C{\sc~iv} and
N{\sc~v} in each component to lie within a factor of 2 of their
observed values and optimizing over the possible sets of
component ionization states and C{\sc~iv} column densities; (2) the
warm absorbing gas exists in the outer part of the wind and is {\it
not} a continuation of the flow in the broad emission-line region; and
(3) the warm absorber extends both in radial and polar directions and
is ionization stratified.  X-ray absorption is found to be heavily
biased towards smaller radii and UV absorption originates at larger
distances from the central continuum source. Our analysis shows that
the discrete absorption components along the line-of-sight are
intrinsically clumpy.  Density differences between kinematic
components result in a range of ionization and recombination time
scales.

We further test the applicability of the MHD wind to warm absorbers in
general, by constructing a quasi-continuous flow model, and extending
it to arbitrary aspect angles. Constraining the ionization and volume
filling factor for a generic case, we estimate the fraction of
Seyfert~1s having detectable warm absorbers with larger O{\sc~vii}
column density than O{\sc~viii}, and the range of total hydrogen column
densities. We also find that the ratio of O{\sc~vii} to O{\sc~viii}
optical depths can serve as a new diagnostic of AGN aspect angle.

Finally, the thermal stability of the UVX absorption model is tested.
We find that all kinematic components in NGC~5548 are thermally stable
to isobaric perturbations. In a more general case, we show that the
magnetic field is crucial in order to stabilize the warm absorber gas
over a wide range of incident continuum spectral energy distribution
and gas metallicity.

\end{abstract}

\keywords{galaxies: active --- galaxies: individual (NGC~5548)
galaxies: nuclei  --- galaxies: Seyfert --- galaxies:  ultraviolet ---
galaxies:  X-rays}

\twocolumn

\section{INTRODUCTION}

Highly-ionized, tenuous, absorbing gas has been detected in the soft
X-ray spectra of many active galactic nuclei (AGNs) at low and moderate
($\Delta E/E = 0.02-0.1$) spectral resolutions.  At least half of the
Seyfert~1 galaxies observed by the \textit{ASCA} satellite indicate the
presence of this ``warm absorber'' gas (hereafter WA) in their X-ray
spectra (e.g., review by Mushotzky 1997).  Other AGNs, such as
radio-quiet and radio-loud quasars, host WA as well (e.g., Green and
Mathur 1996, Mathur and Elvis 1995, Siebert et al.\ 1996, Ulrich et
al.  1999), though no statistics are currently available.  In addition,
broad absorption-line quasars (hereafter BAL~QSOs), which constitute at
least 10\% of all quasars, are conspicuously weak in the soft X-rays
--- a possible signature of a WA gas (Mathur, Elvis, \& Singh 1995;
Green \& Mathur 1996; Gallagher et al.\ 1999).

The X-ray WA gas is inferred from the softening of the continuum at the
low X-ray energies, between $0.6$ keV and a few keV, and is due to many
unresolved absorption edges from highly ionized species of elements,
such as O, Ne, etc. The strongest edge depths correspond to O{\sc~vii}
and O{\sc~viii} at 0.72 and 0.85 keV, respectively.  It is generally
found that the total hydrogen column density of absorbing material lies
between $10^{21}\ {\rm cm^{-2}}$ and $10^{23}\ {\rm cm^{-2}}$, and may
even reach $10^{24}\ {\rm cm^{-2}}$ (Reynolds et al.\ 1997, hereafter
R97; George et al.\ 1998, hereafter G98). The energy resolution of the
\textit{ASCA} satellite data does not allow for a study of the
kinematics of the WA gas based on the X-ray absorption profiles.  The
thermodynamic state of the gas, however, including its temperature and
density, can be inferred with a moderate amount of modeling and
invoking observed X-ray variability (Otani et al.\ 1995, Reynolds \&
Fabian 1995; R97). 

Intrinsic blueshifted UV absorption was detected in AGNs by the {\em
IUE} (Bromage et al.\ 1985; Koratkar et al.\ 1996; Leech et al.\ 1991;
Walter et al.\ 1990; Voit et al.\ 1987; Ver\'{o}n et al.\ 1985) and
{\em HUT} (e.g., Kriss et al.\ 1992). High and sometimes low-ionization
lines are observed, e.g., C{\sc~iv}, N{\sc~v}, O{\sc~vi},
Si{\sc~iv}, Mg{\sc~ii}, and others. The higher S/N and spectral
resolution data taken by the {\em HST} found many more cases of
blueshifted intrinsic absorption in the UV resonance lines. Preliminary
statistics indicate that intrinsic UV absorption is present in at least
half of Seyfert~1 galaxy spectra (Crenshaw et al.\ 1999 [C99]).  In
addition to the BAL~QSOs, there are some quasars that exhibit intrinsic
narrow absorption features (see, for example, Mathur, Wilkes, \& Elvis
1999).

In all known cases, the UV resonance line absorption is present in
sources with known soft X-ray absorption, hinting of a connection
between the two types of absorption (C99; Mathur et al.\ 1999;
Brandt, Laor, \& Wills 1999), though the nature of this connection is
not yet clear.  The correspondence between the UV and X-ray absorbing
gas was advanced by Mathur et al.\ (1995), on the basis of single slab
modeling of UVX absorbing gas in NGC~5548, using {\it HST/FOS}, {\it
ROSAT}, and {\em ASCA} data in their analysis.  Similar analysis was
performed for other objects (e.g., NGC~3516, Mathur et al.\ 1997;
NGC~3783, Shields \& Hamann 1997).  The mean FOS spectrum of NGC~5548
represented the best UV data at the time, and its low spectral
resolution ($\sim 200~{\rm km\ s^{-1}}$) justified the single slab
approximation. Much higher spectral resolution {\em HST} GHRS and
STIS/Echelle have since revealed the presence of a number of distinct
kinematic components within the outflowing material, with FWHM of
$50~\kms - 160~\kms$ (Crenshaw \& Kraemer 1999).  These high resolution
spectra of NGC~5548 have uncovered at least five separate C{\sc~iv}
absorption components (C99; Crenshaw \& Kraemer 1999; Mathur,
Elvis, \& Wilkes 1999 [MEW99]).

The presence of a structure within the intrinsic absorption adds
another degree of complexity to the kinematical models of AGNs.  It is
important, therefore, to understand the relationship between the
outflowing highly-ionized gas, the broad emission-line (BEL) and
absorption-line regions in these objects. Numerous models have been
proposed to address this problem (e.g., Blandford 1990; Arav, Shlosman
\& Weymann 1997; K\"onigl \& Kartje 1994). In this paper we extend the
hydromagnetic wind model of Emmering, Blandford and Shlosman (1992,
hereafter EBS), which provides a dynamical explanation for the the
properties of the broad emission-line regions (BLRs) in AGNs (Bottorff
et al.\ 1997, hereafter Paper~1), to the UVX absorption phenomena. In
the EBS model, gas in a dusty molecular accretion disk is loaded on
magnetic field lines and centrifugally launched forming a magnetized
wind. The basic difference between the EBS model and other outflow
models is that it provides for a 3D wind dynamics, with helical motion
around the axis being superposed onto the radial motion. The wind is
clumpy and stratified both in density and ionization. Paper~1 addressed
the radial wind structure at smaller latitudes above the
optically-thick disk. First, it was found that the BLR is geometrically
thick and has substantial optical depth.  As a result, the inner part
of the wind becomes progressively ionized with increasing latitude,
while the outer part of the wind remains cold and molecular as long as
it is shielded by the BLR. As the outer
material continues to rise it becomes ionized and eventually passes the
observer's line-of-sight, where the wind is detected via absorption in
both UV and X-ray energy bands.  The line-of-sight velocity of the wind
depends on its injection radius on the disk, so a range of velocities
is expected.  The hydromagnetic wind model, therefore, naturally
explains both the UV emission and absorption systems seen in C{\sc~iv}
and other lines, as well as X-ray absorption. While our approach is
general, we emphasize the results within the context of observations of
NGC~5548.

In Section~2, we summarize the important observations of the WA gas,
including the column densities of O{\sc~vii}, O{\sc~viii} and total H
from the X-ray data, and the column densities of C{\sc~iv}, N{\sc~v}
and H{\sc~i} from the UV data.  In Section~3, we consider specifically
the case of NGC~5548 and its discrete UV absorption components.  The
velocity field of the EBS model adopted in Paper~1 is then used to
determine the location and line-of-sight width of the UV absorption
components.  The observed column densities of N{\sc~v} and C{\sc~iv} in
conjunction with the location and line-of-sight width of each component
allow us to determine the density, temperature, volume filling fraction
and column densities of O{\sc~vii}, O{\sc~viii}, H{\sc~i} and total H
via photoionization modeling.  The calculated ionic column densities
are compared with observations, to determine whether or not the X-ray
absorption in NGC~5548 is compatible with the model and in order to
determine the association between the UV and X-ray warm absorbers.

Section~4 describes a generic WA model. The multicomponent absorption,
as seen in the C{\sc~iv} UV absorber of NGC~5548, is smoothed into an 
equivalent single
absorbing column, to study the effects of a distributed WA in Seyfert
galaxies without specific knowledge of kinematic absorption
components.  The generic model is investigated for arbitrary aspect
angles to the observers.  We estimate the fraction of AGNs having
detectable WA gas, and larger O{\sc~viii} than O{\sc~vii} optical
depth.  Finally, in the Appendix, we discuss the general thermal
stability of magnetized UVX absorbing gas.

\section{OBSERVATIONAL SUMMARY}
 
The spectral resolution of \textit{ASCA} is insufficient to clearly
resolve and measure oxygen and other element absorption edges without
additional modeling. This is achieved by producing simplified
photoionization models of WA gas using observed AGN continua.  Free
parameters in the model are subsequently tuned to obtain the best fit
for the calculated transmitted spectrum to the observed one.

A common approach taken by R97 and G98 makes a number of essential
assumptions, namely, the WA is considered (1) to be a thin slab; (2) to
have a constant total hydrogen number density; (3) to be in thermal
equilibrium; (4) to be exposed to a power-law soft X-ray continuum of
photon index $\Gamma \sim 2$, and (5) to possess solar abundances.  The
total hydrogen column density (or equivalently slab thickness) $N_{H}$,
the X-ray ionization parameter $U_{X}$ and $\Gamma$ are then varied
until an optimal fit to the data is obtained.  Additional free
parameters include Galactic absorption, a fraction of unattenuated
central continuum, a covering factor of the emitting gas, etc.

The main inferred properties of WA gas from X-ray observations are the
optical depths of O{\sc~vii} and O{\sc~viii}, and the column density of
total H along the line-of-sight.  Ionization stratification in WA gas
due to differences in density and location is indicated in at least one
object, namely MCG~6-30-15. The latter exhibits variability in the
O{\sc~viii} edge depth which is anticorrelated with the continuum, and
little or no variability in O{\sc~vii} (Fabian et al.\ 1994, 1995;
Otani et al.\ 1995).  The difference in response is frequently attributed 
in literature to the
recombination time scale, $t({X_i})$, of ion $X_i$ given by
\begin{equation}
t(X_i)=1/\alpha(X_i,T_e)n_e,
\end{equation}    
where $n_{e}$ is the electron density, and $\alpha(X_i,T_e)$ is the
recombination coefficient, which is a function of the ion species $X_i$
and a slowly varying function of electron temperature $T_{e}$ (but see
section 3.3.5).  In such a case, over a
wide range in temperature $\alpha(O{\sc~vii},T_e)$ and
$\alpha(O{\sc~viii},T_e)$ are of the same order of magnitude, so
recombination time differences between the two ions are mainly due to
differences in density.  The response of O{\sc~viii} and the lack of
response of O{\sc~vii} in MCG~6-30-15 during the observing time is
interpreted as evidence that O{\sc~vii} absorption occurs in a lower
density gas than the O{\sc~viii} absorption.  Photoionization modeling
and absorption depth variability in MCG~6-30-15 also indicate that the
O{\sc~vii} and O{\sc~viii} absorption occurs at different locations
(Otani et al.\ 1995). An additional indication of ionization
stratification in WA AGN gas is that objects with significant optical
reddening exhibit largest O{\sc~vii} absorption --- a hint that X-ray
absorption may occur in regions where dust can survive (Komossa \& Fink
1997).

If the UV and X-ray absorbers are related, then some of the assumptions
used in previous modeling efforts must be modified in view of the
detection of multiple kinematic components in Seyfert galaxies, in
particular in NGC~5548, which exhibits 5 separate, blueshifted UV
absorption components. Thus, we abandon the single zone model in favor
of a multiple zone constrained by the EBS model.  The constant density
approximation is relaxed by estimating density for each of the
kinematic components separately, by placing them at different distances
from the central continuum source, based on kinematical constraints of
the EBS model. We assume thermal and ionization equilibrium in the UVX
gas (but see Krolik \& Kriss 1995, and Nicastro et al.\ 1999).  Target
values for the total hydrogen column density, $N_{H}$, and the optical
depths $\tau(O{\sc~vii})$ and $\tau(O{\sc~viii})$, which we convert to
column densities, are adopted from the X-ray observations discussed in
R97 and are listed in Table~1a.

In addition to the observations at X-ray energies, we also consider the
{\em HST} UV GHRS observations of the 5 discrete, blueshifted
absorption components of C{\sc~iv}, N{\sc~v} and H{\sc~i} detected in
NGC~5548 on 1996 Aug.~24 and on 1996 Feb.~17
(C99) (see Table~1b). We adopt the ionic column densities
measured from the GHRS observations over the STIS observations
(Crenshaw \& Kraemer 1999) because the former are closer in time to
the {\em  ASCA} X-ray observations of NGC~5548 on 1993 July~27 (R97).
The single exception is for the column density of $H{\sc~i}$ which is 
lacking in component 1 in the GHRS spectra so we adopt the value from
the STIS data. 
Note that the UV observations are separated by seven months and
the X-ray observations are separated from the UV observations by about
2 years. Extensive multiwavelength monitoring campaigns of this object
(Clavel et al.\ 1991; Korista et al.\ 1995) show variations of UV
emission lines which correlate with the continuum over times of days to
weeks.  C{\sc~iv} emission in NGC~5548 has, for example, varied in
strength by a factor of $\sim 2$ in a month's time and C{\sc~iv}
absorption equivalent width may be similarly variable (Shull \& Sachs
1993).  Caution must therefore be taken when using noncontemporaneous
data to determine the ionization state of the absorbing gas.  We also
point out that reported values of the ionic column density depend upon
the procedure used to fit absorption line profiles.  For example, given
the same set of {\em HST}/GHRS spectra of the C{\sc~iv} absorption
systems, MEW99 and C99 derived values of the C{\sc~iv} column
densities of components $1-3$ which differed by $0.2-0.5$~dex.  Next,
modeling that goes into determining observationally-derived ionic
column densities does not explicitly account for the possible effects
of continuum scattering into the absorption troughs, although these
effects do appear grossly in the estimate of the incomplete
line-of-sight continuum coverage.  This modeling must also make
assumptions concerning the underlying emission from the narrow and BEL
regions.  Finally, the absorption lines are often observed to be
saturated, yet non-black. The line-of-sight covering fraction of the
ions along the various line-of-sight velocities must be estimated from
the degree of saturation present.  This is usually done by comparing
the depths of the resonance line doublet pairs.  Differences in
line-of-sight effective continuum coverage amongst the various ions can
be expected and are often found. The H{\sc~i} column densities reported
by C99 and Crenshaw \& Kraemer (1999) suffer from a lack of
constraints in this respect (Ly$\alpha$ is a singlet, and no higher
order Lyman line data are available), and should therefore be
considered as lower bounds (Crenshaw, private communication).
Thus the uncertainties in the ionic column densities reported in the
literature are internal uncertainties only, the total uncertainties
must be significantly larger.

Our adopted ionic column density set is listed in Tables~1a and
1b.  Given the above considerations, we have disregarded their
reported uncertainties, and instead assigned a factor of 2 uncertainty
above and below their reported values. An exception is component~1 of
C{\sc~iv}, which at the limit of detection is treated as a lower bound
(Crenshaw, private communication). These assigned ranges of uncertainty
about the reported ionic column density values, which we will refer to
as observational constraints, will be used to find a {\it plausible}
set of solutions to the UVX absorber in NGC~5548.

\section{DYNAMICS AND EMISSION MODELING FOR AGNs}

\subsection{Hydromagnetic Wind Model}

The magnetohydrodynamic (MHD) solution for a stationary, axisymmetric
flow in cylindrical coordinates is given by Blandford \& Payne (1982)
and describes a self-similar, cold, nonrelativistic MHD flow from an
idealized Keplerian disk around a point mass.  This solution can be
written in terms of variables $\chi$, $\xi$, $\phi$ and a scaling
parameter $r_0$ which are related to cylindrical coordinates via ${\bf
r}\equiv [r_0\xi(\chi), \phi, r_0\chi]$.  Here, $\chi$ is the
coordinate along a field line, $\xi(\chi)$ is found as part of a
self-consistent solution to the MHD equations, and $r_0$ is the field
line footpoint on the disk.  The flow velocity components are given by
\begin{equation}
{\bf v} =[\xi'(\chi)f(\chi),g(\chi),f(\chi)]\sqrt{GM/r_0}, 
\end{equation}
where a prime denotes differentiation with respect to $\chi$, and $M$
is the mass of the central black hole.  The EBS
provide for a non-trivial extension of Blandford \& Payne solution for an
arbitrary scaling of volume density $n\propto r_0^{-b}$ and magnetic field
$B\propto r_0^{-(b+1)/2}$.  At the base of the flow, the rotational
velocity, $v_\phi$, is Keplerian so $v_\phi\propto r_0^{-1/2}$.  The
functions $\xi(\chi)$, $f(\chi)$ and $g(\chi)$ are chosen to satisfy
the flow MHD equations subject to the above scalings of $\rho$, $B$ and
$v_\phi$.  This implies that the Alfv\'{e}n speed scales with the disk
Keplerian velocity, and the specific angular momentum and energy in the
flow will scale similarly to their Keplerian counterparts, while the
disk mass loss per decade in radius scales as $\propto r_0^{-(b-1.5)}$. The
above velocity law is of course a non-Keplerian one at all finite distances
from the disk  (e.g., EBS and Paper~1). 

EBS utilized an analytic approximation for $\xi(\chi)$ in which
$\xi(\chi) \sim \chi^{1/2}$, and asymptotically tends to the Blandford
\& Payne solution.  To attach the solution to the disk at $r_0$,
$\xi(\chi)$ was constructed to have the form $\xi(\chi) =
\sqrt{\chi/c_2+1}$, so that $\xi(0)=1$.  The constant $c_2=({\rm
tan}~\theta_0)/2$, where $\theta_0$ is the initial angle between the
magnetic line and the disk.  The MHD equations have been solved for
$f(\chi)$ and $g(\chi)$, so as to be consistent with the analytic form
of $\xi(\chi)$. The model parameters have been fixed in Paper~1 by modeling
the BLR response to the variation of the central continuum (see section~3.2). 
We use $b=1.5$, so $n$, the particle density, scales as $n\propto
R^{-3/2}$ and $B\propto R^{-5/4}$, where $R$ is the spherical radius from the
central mass. We show below that this choice in $b$ is consistent with the
variation in density inferred from observations, although the number of
observational points is small. This result, therefore, should be taken with
necessary caution. We have further
neglected the effect of radiation pressure on the gas dynamics (e.g., de Kool
\& Begelman 1995), as it will not qialitatively change our results.

\subsection{Constraining the MHD Flow with the BLR Spectrum}

With an analytical solution for the volume emissivity EBS showed that
realistic BLR emission line profiles can be produced.  Paper~1
continued these efforts replacing the analytic emissivity function with
an emissivity function obtained from a fit to an amalgam of optically
thick clouds calculated by the photoionization code Cloudy (Ferland
1996).  Emission line anisotropy and finite optical depth effects were
introduced to model C{\sc~iv} emission from optically thick BLR clouds
in the well studied object, NGC~5548.  As part of the modeling,
variations in C{\sc~iv} emission in response to observed variations of
the continuum, including light travel time effects, were studied.  A
time series of synthetic C{\sc~iv} emission line profiles were
generated and compared with the observed C{\sc~iv} emission line
profiles from the {\em HST} (Korista et al.\ 1995), and a model fit to
the data was thereby achieved.  The BLR parameters in NGC~5548 deduced
from the model were the central black hole mass ($3\times10^{7}
M_{\odot}$), physical extent of the C{\sc~iv} emitting gas (1 to 24
light days) and the orientation of the observer line-of-sight relative
to the axis of symmetry ($40^\circ$).  Here, we extend the results of
Paper~1 and analyze the flow beyond the BLR and at larger latitudes
above the disk, in particular as the flow crosses the observer's
line-of-sight.

In the ideal EBS flow, cold molecular gas is launched from a Keplerian
disk and is flung along the magnetic field lines, like beads along a
rotating wire.  Most of the magnetic lines, however, are inclined
unfavorably and hence will not be able to accelerate the gas.  The
centrifugally accelerated gas becomes illuminated by the central
ionizing continuum at some latitude.  The distribution of densities
and ionization parameters, combined with the MHD model's velocity field
produces typically observed broad emission-line profiles.  For
NGC~5548, Paper~1 found that the BLR lies within a toroidal wedge $\pm
30^\circ$ of the equatorial disk plane.  The boundary of the wedge was
interpreted as coming from optically-thick wind filaments experiencing
thick-to-thin transition. In this section, we focus on the optically
thin (to the ionizing continuum) flow, which is the continuation of the
BLR flow, as it crosses the line-of-sight to the observer. Due to 
rather unfavorable conditions for loading and acceleration at the base
of the wind, it is natural to expect a number of separate kinematic
components along the line-of-sight. The five blueshifted UV
absorbers reported in C99 and MEW99 for NGC~5548, while perhaps not
fully distinct from one another, are interpreted here as kinematically
distinct regions containing gas in an EBS-type magnetized wind.
Moreover, each individual component is expected itself to be clumpy and
is characterized by a volume filling factor, as we show below.

A point on a streamline is defined in terms of the footpoint radius
$r_0$, the spherical radius $R$, and the observer's aspect angle $i$
with respect to the disk axis, namely
\begin{equation}
r_{0}=Q(i,\theta_0)R, 
\end{equation}
where 
\begin{equation}
Q(i,\theta_0)=\frac{\sqrt{{\rm cos}^2i+{\rm tan}^2\theta_{0}{\rm sin}^2i}
   -{\rm cos}i}{{\rm tan}\theta_{0}}. 
\end{equation}
\begin{figure}[ht]
\vbox to3.6in{\rule{0pt}{3.6in}} 
\includegraphics{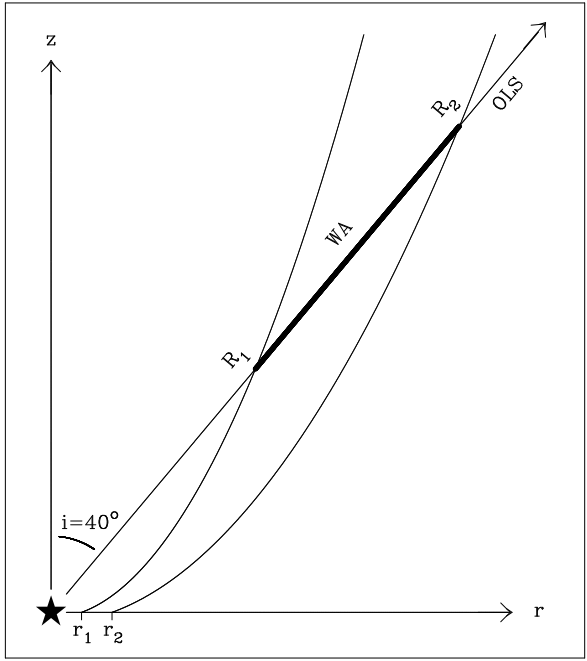} 
\caption{Schematic representation of the warm absorber
column.  Gas, destined to contribute to the warm absorber, rises off a
disk in the $z=0$ plane between $r_{1}$ and $r_{2}$ and crosses the
observer's line-of-sight (OLS) between $R_{1}$ and $R_{2}$, where it
is detected via continuum absorption.  Gas inside
$R_{1}$ is overionized, so there is little contribution to the
absorption column densities of O{\sc~vii}, O{\sc~viii}, C{\sc~iv} or
N{\sc~v} in this region.  Outside $R_{2}$ the flow has been cut off by
mechanisms described in the text.}
\end{figure}  
In Paper~1, for the best fit model of the BLR in NGC~5548, the launch
angle $\theta_0$ and aspect angle $i$, were found to be $20^\circ$ and
$40^\circ$, respectively. In this case, $R\approx10.4 r_0$. Figure~1
illustrates the model geometry for NGC~5548.  A patch of gas filaments
lifts off the disk between cylindrical radii $r_1$ and $r_2$ and rises
along a helical trajectory, forming a parabola in the $rz$-plane.  Gas
that starts at $r_1$ eventually crosses the observer line-of-sight at a
spherical radius $R_1$ and gas that starts at $r_2$ does this at
$R_2$.

The number density of a total hydrogen along the flow line that starts
at $r_0$ is prescribed by the model to be
\begin{equation}
n=\frac{n_{A}}{m}\left(\frac{r_0}{r_1}\right)^{-1.5}.
\end{equation}
Here $n_A$ is a normalization constant, setting the density $n$ on the
innermost flow line at $m=1.0$, where $m$ is the Alfv\'{e}n Mach
number.  The value of $m$ increases from unity at about $17^\circ$
above the disk to about $10^{3.3}$ at $85^\circ$ above the disk. The
ionization parameter $U\propto 1/nR^2$, therefore, scales as $U\propto
mQ^2(i,\theta_0)$ along a flow line. At small latitudes, the
effective ionization parameter for lower column density filaments will
be even smaller, due to finite optical depth effects in the BLR
(Paper~1).  The change in radius and density causes the ionization
parameter along any flow line to have a maximum at about $30^\circ$
above the disk, where the BLR flow becomes optically thin. Here, the BLR 
flow is no longer 
self-shielded, becomes overionized and remains so
at larger latitudes, where it crosses the observer's
line-of-sight.  As a result, this gas does not contribute significantly
to the UVX absorbing column. The gas filaments launched at larger
radii than the BLR flow are less ionized and form the bulk of the UVX
absorption. This absorbing gas is indicated by the thick line segment
in Figure~1.

While the inner boundary of the wind is set by photoionization, the
outer boundary can be fixed by considering the amplification of
magnetic fields in the disk.  An efficient generation of magnetic
fields in the disk will cause its expulsion through the buoyancy
effects and probable field reconnection above the disk (e.g., Galeev,
Rosner \& Vaiana 1979).  The energy release during the reconnection can
in principle inject disk material and load it onto the large-scale
field treading the disk, resulting in the hydromagnetic wind.  While
the large-scale disk-threading poloidal field is probably governed by
global considerations, the disk-generated fields are expected to be
amplified by differential rotation.  Such dynamo effect will operate
only within the ``radius of influence,'' $r_{BH}$, of the central black
hole (BH),
because for $r > r_{BH}$ the inner galactic rotation is typically that
of a solid body.  Hence, the outer wind cutoff can plausibly result
from a dramatic decrease in the shear outside $r_{BH}$, which prevents
the magnetic field in the disk from reconnecting and loading the
molecular material onto large-scale field lines.  This radius can be
estimated from equating the gravitational potentials of the BH and that
of the host galaxy, namely, $r_{BH}\sim 1.3\times 10^{19} m_7
v_2^{-2}$~cm, where $m_7\equiv M_{BH}/10^7\ \msun$ and $v_2 \equiv
v_K/100\ \kms$ is the rotational velocity in the galactic disk.
Adopting the BH mass in NGC~5548 from Paper~1, leads to $R_2\sim
1.4\times 10^{20}$~cm for a typical $v_7=2.5$.  As we show below in
$\S$~3.3.1 this is within a factor of $\sim 2-3$ from the outer
boundary of the observed kinematic components.

\subsection{Multi-Component Absorption Model for NGC~5548}

\subsubsection{Location of Absorption Components}

The positions of separate absorption components are calculated using
the relationship between $R$ and $v_{obs}$, the line-of-sight component
of wind velocity.  The velocity $v_{obs}$ is obtained by projecting the
velocity field onto the observer line-of-sight and is given by
\begin{equation}
v_{obs} =[\xi'(\chi)f(\chi){\rm sin}i+f(\chi){\rm cos}i]\sqrt{GM/r_{0}}.
\end{equation}
Substituting $r_{0}=Q(40^\circ,20^\circ)R$ and $M=3\times10^{7}\ \msun$ 
for NGC~5548 (Paper~1), gives
\begin{equation}
R = 7.3\times 10^{18} \left(v_{obs}\over 10^3~\kms\right)^{-2}\ {\rm cm}. 
\end{equation}
Observed component velocities in the rest frame of NGC~5548 from C99 and the
FWHM of each component which we use below are listed in Table~3b.  Inserting the
observed component velocities into Equation~7, yield component distances which
range from about 2 to about 87 parsecs (Table~3b).  The most important
consequence is that the UVX gas in NGC~5548 appears to be positioned well
outside its BLR, both in radii and in latitude.  Hence, within the framework of
the EBS model, which neglects the effects of radiation pressure, the the UVX
absorbing gas originates in the outer part of the wind, beyond the BLR flow. 

\subsubsection{Calculation of Possible Ionization States}

The observational constraints on the ionic column densities and their
ratios in section~2 are now used to find the limits on the ratios of
the N{\sc~v} ion fraction, $f(N{\sc~v})$, with the C{\sc~iv} ion
fraction, $f(C{\sc~iv})$.  These estimates allow the determination of a
set of possible gas ionization states, i.e., the particle density,
electron temperature and various ion fractions in each of the UVX
absorption components.  We assume a total hydrogen number density $n$
as given by equation~5, so the column density $N(X_i)$ of ion $X_i$ is
\begin{equation}
N(X_i)=\int_{\Delta R}^{ }\epsilon f(X_i)a_{X}ndR \approx \epsilon f(X_i)a_{X} 
\int_{\Delta R}^{ }ndR,
\end{equation}
where $\epsilon$ is the volume filling factor in the gas column which
covers the central continuum source, $f(X_{i})$ is the ion fraction,
$a_X$ is the abundance of element $X$, and $\Delta R$ is the
line-of-sight geometrical width of the cloud.  If the ionization
parameter is a slowly varying function of radius with little change
over $\Delta R$ and $\epsilon$ is constant within each component, then
$\epsilon f(X_i)a_{X}$ can be taken out of the integral.  Then
$f(N{\sc~v})/f(C{\sc~iv})$ for each component is given by
\begin{equation}
\frac{N(N{\sc~v})}{N(C{\sc~iv})} \approx 
\frac{f(N{\sc~v})}{f(C{\sc~iv})}\frac{a_N}{a_C}.
\end{equation}
Assuming solar abundances, $a_N=9.33\times 10^{-5}$ and
$a_C=3.55\times 10^{-4}$, we obtain the ion fraction ratio
\begin{equation}
\frac{f(N{\sc~v})}{f(C{\sc~iv})} \approx 3.805 \frac{N(N{\sc~v})}{N(C{\sc~iv})}. 
\end{equation}

The central values of $\log[f(N{\sc~v})/f(C{\sc~iv})]$ for components
$1-5$, assuming the column densities listed in C99 and solar
abundances, are listed in the final column of Table~1.  In accord with
section~2, we assign factors of two uncertainty to their values.
Component~1 is a special case because the reported values for
$N(C{\sc~iv})$ provide a lower bound only.

The range of possible ionization states for each component was
determined using Cloudy (version 90.04).  The luminosity and spectral
energy distribution (SED) of the continuum source in NGC~5548, combined
with the distance of each component (Eq.~7), is used to calculate the
incident flux $\Phi(H)$ of hydrogen ionizing photons falling on each
absorption component.  We used the following continuum shape for the
SED of NGC~5548, i.e.,
\begin{equation}
f_{\nu} \propto 
\nu^{\alpha_{UV}}e^{-h\nu/kT_{BB}}e^{-kT_{IR}/h\nu}+a\nu^{\alpha_{X}} 
\end{equation}
and
\begin{equation}
\frac{f_{\nu}(2~{\rm keV})}{f_{\nu}(2500~{\rm \AA}\/)}=403.3^{\alpha_{OX}},
\end{equation}
where $kT_{IR}=0.01$~Ryd is the spectrum cutoff in the near-infrared,
and $a$ in Equation~11 is chosen to satisfy
Equation~12.\footnote{Outside the range 1.36~eV$ - $100~keV, $a=0$.
Above 100~keV, we assume $f_{\nu}\propto\nu^{-3}$.} The input
parameters are ${\rm log}~(T_{BB})=6.683$, $\alpha_{OX}=-1.20$,
$\alpha_{UV}=-1.20$ and $\alpha_{X}=-0.90$, chosen to be in accord with
{\it IUE/ROSAT} observations of the NGC~5548 continuum (Walter et
al.\ 1994). Using the above SED, the ionizing luminosity of the
continuum source is estimated at $10^{44.3}~\rm{erg~s^{-1}}$
($H_0=75~\kms~{\rm Mpc^{-1}}$), based upon the mean continuum flux at
1350\AA, obtained during the 1993 observing campaign (Korista et
al.\ 1995).

With the above assumptions, the ionization parameter $U$
($\equiv\Phi[H]/nc$), which controls the ionization state of the gas,
is tuned by adjusting the hydrogen number density $n$ in a series of
Cloudy calculations for an optically-thin slab of width $10^{10}\rm
cm$.  In this manner, a set of synthetic photoionization gas states
covering a range of possible $f(N{\sc~v})/f(C{\sc~iv})$ values is
produced for each component.

In order to provide for a complete description of the UVX absorption in
each kinematic component, it is necessary to know the column densities
of different ions. Note that, if we replace $N(N{\sc~v})$ with
$N(X_{i})$ and $a_{N}$ with $a_{X}$ in Equation~9, we obtain a generic
expression for the column density of ion $X_{i}$.  In the approximation
of Equation~9, the quantity $[f(X_{i})/f(C{\sc~iv})](a_{X}/a_{C})$ is
constant, so $N(X_{i}) \propto N(C{\sc~iv})$.  Thus for fixed
$f(N{\sc~v})/f(C{\sc~iv})$, all ionic column densities scale directly
with the C{\sc~iv} column density.  The parameters
$f(N{\sc~v})/f(C{\sc~iv})$ and $N(C{\sc~iv})$ then map out the set of
possible ionic column densities for each component.  We use these two
parameters to determine the electron temperature $T_{e}$ and column
densities of N{\sc~v}, O{\sc~vi}, O{\sc~vii}, O{\sc~viii}, H{\sc~i} and
H, on a grid of $[$f(N{\sc~v})/f(C{\sc~iv})$,N(C{\sc~iv})]$ pairs that
are consistent with the observational constraints.

The grid of solutions constrained by f(N{\sc~v})/f(C{\sc~iv}) and
N(C{\sc~iv}) were used to determine the range of possible values of the
column densities of O{\sc~vi}, O{\sc~vii}, and O{\sc~viii} in each
component.  The factor of 2 uncertainty in the observational
constraints on the C{\sc~iv} and N{\sc~v} column densities and their
ratios translates into column density uncertainties of a factor of
$\sim 10$ on O{\sc~vi}, a factor of $\sim 100$ on O{\sc~vii} and a
factor of $\sim 1,000$ on O{\sc~viii}. The inferred ranges on the
column densities of H and H{\sc~i} are no better. H{\sc~i} ranges over
a factor of $\sim100$ and H over a factor of $\sim 1,000$ (even more in
component~1).  We, therefore, obtain an optimal solution for UVX
absorption in NGC~5548, within the five component sets of possible
solutions. Since individual column densities of C{\sc~iv} and N{\sc~v}
for each component are observed, but only total column densities are
available for O{\sc~vii} and O{\sc~viii}, we search for a solution that
minimizes the chi-squares, $\chi_{O}^{2}$, for oxygen ion column
densities, namely
%
$$\chi_{O}^{2} =
\frac{(\sum_{j=1}^{5}N(O{\sc~vii})_{j}-N(O{\sc~vii})_{obs})^{2}}              
          {\sigma_{O{\sc~vii}}^2} $$
\begin{equation}
+ \frac{(\sum_{j=1}^{5}N(O{\sc~viii})_{j}-N(O{\sc~viii})_{obs})^{2}}
                        {\sigma_{O{\sc~viii}}^2}. 
\end{equation}
Here $N(\mathrm{O{\sc~vii}})_{j}$ and $N(\mathrm{O{\sc~viii}})_{j}$ are
the calculated column densities of O{\sc~vii} and O{\sc~viii} in a
component $j$ (see Table 3), $N(\mathrm{O{\sc~vii}})_{obs}$ and
$N(\mathrm{O{\sc~viii}})_{obs}$ are the observed total column densities
of O{\sc~vii} and O{\sc~viii} (see Table~1), and $\sigma_{O{\sc~vii}}$
and $\sigma_{O{\sc~viii}}$ are the adopted observational uncertainties
(section 2) in the O{\sc~vii} and O{\sc~viii} column densities.  Our
search of the solution grid yields optimal $\chi_{O}^{2} = 0.13$. The
total column density of O{\sc~vii} is overpredicted by $30\%$ and the
total column density of $O{\sc~viii}$ is underpredicted by $9\%$,
compared to target values in Table~1a, placing both well within the
observational constraints.

\subsubsection{An Optimal Solution for UVX Absorption using the EBS Model}

The model column densities of C{\sc~iv}, N{\sc~v}, O{\sc~vi},
O{\sc~vii}, O{\sc~viii}, H{\sc~i} and H in each component and their sum
over all components are listed in Table~3a.  Component C{\sc~iv} and
N{\sc~v} column densities automatically satisfy the observational
constraints because the optimization was done on a grid of values
bounded by the constraints. The component column density values for
O{\sc~vi}, O{\sc~vii}, O{\sc~viii}, and H in each component are
predictions of our model. The good match between the predicted and
observed total O{\sc~vii}, O{\sc~viii}, and hydrogen column densities,
given present observational constraints, indicates that the soft X-ray
WA and the UV resonance-line absorber are {\it directly} related.  The
EBS model constrains the bulk of the WA to lie within two of the
kinematic ``components'' of the outflow, separated by about $500\ {\rm
km\ s^{-1}}$ in line-of-sight velocity. While observations at the
required resolution do not yet exist to test these predictions in
detail, the recently launched X-ray satellite {\em Chandra}  may be
able to detect the O{\sc~vii} and O{\sc~viii} absorption edges of the
individual kinematic absorption components in NGC~5548, and the
recently launched {\em FUSE} satellite should at least detect the
presence of absorption in O{\sc~vi}.

The predicted column densities of H{\sc~i} from our optimal solution do
not match well their reported values (Table~1b). The observed component
column density values are factors of $4-28$ below their counterparts in
the optimal solution. However, as mentioned in section~2, the
observationally inferred H~I column densities are the least
certain, and are lower limits to their
true values, at best.

The hydrogen number density for each component in the optimal solution
as well as $T_{e}$, $U$ and other physical values are listed in
Table~3b.
The density derived for each component using Cloudy depends on the kinematic
structure of the EBS model via Equation~7, but is \textit{independent} of the
parameter $b=1.5$ used to describe the scaling of the EBS density in Equation~5.
It is therefore possible to test whether the assumed value of $b$ used in Paper~1
and here in this paper is appropriate.

\begin{figure}[ht] 
\vbox to2.6in{\rule{0pt}{2.6in}} 
\includegraphics{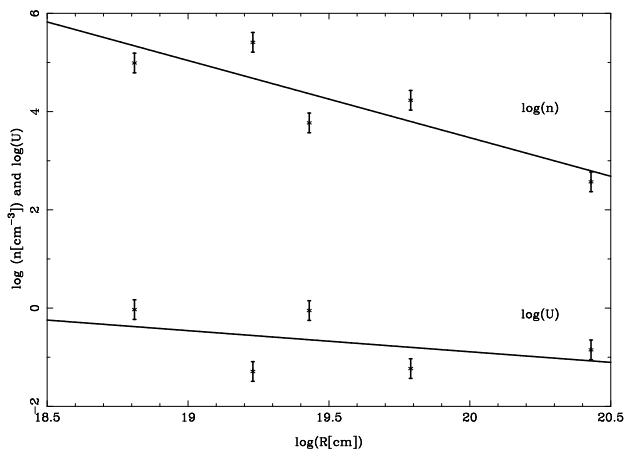}    
\caption{Logarithmic plot of density
and ionization parameter as a function of $R \rm{[cm]}$ for the discrete
component model of NGC~5548. The straight lines are the least squares fits.}
\end{figure}

Equations~3--5 together imply that along any ray through the origin
$n \propto R^{-1.5}$, and, therefore, the ionization parameter varies
as $U\propto R^{-0.5}$.  If individual absorption components obey this
density prescription, then $n$ and $U$ from component to component
should vary accordingly.  As a test, we fit $\log(n)$ and $\log(U)$
from the optimal solution at the center of each absorption component,
as functions of $\log(R)$ with a linear least squares fit.  Figure~2
shows $\log(n)$, $\log(U)$ and the least squares fit of both.  The
range in uncertainty is $\pm 0.2$ and corresponds to our
photoionization grid's resolution in gas density.  We find that the
linear least squares fits to $\log(n)$ and $\log(U)$ {\em vs.}  $\log
(R)$ have slopes of $-1.57\pm 0.16$ and $-0.45\pm 0.16$, respectively.
These are consistent with our specific EBS assignment of $n \propto
R^{-1.5}$ and $U \propto R^{-0.5}$.  Since $U$ is a slowly decreasing
function of $R$, our simplification of the integral in Equation~8 is
reasonable, assuming that there are no significant optical depth
effects along the absorbing column and that in each component $\Delta
R/R$ is sufficiently small, so that there are no significant changes in
$U$ over the width of a component either.  The picture that emerges for
the wind model of NGC~5548 is that of {\it optically thin} filaments
originating over a large range of radii \textit{in the disk}, from
$\sim 0.2$~pc to about 9~pc.  These filaments being centrifugally
accelerated along the magnetic field lines, cross the line-of-sight
about ten times further out.  Along the line-of-sight, the ionization
parameter drops as a weak function of distance ($U\propto R^{-1/2}$),
but because the components are spread out over a factor of $\sim 40$ in
radius, there is a factor of $\sim 6$ drop in ionization parameter so
the wind is ionization-stratified along the line-of-sight.  The thermal
stability of kinematic components in NGC~5548 is analyzed in the
Appendix.

\subsubsection{Absorption Component Widths and Volume Filling Factors}

A more complete physical picture of the UVX absorption system is
provided by determining the line-of-sight width, $\Delta R$, of each
absorption component. In this section we estimate the ratio $\Delta
R/R$, discuss the radial gas distribution, separation of individual
components, their degree of clumpiness, variation of $U$, etc.

The line-of-sight extension of each kinematic component can be
estimated from the FWHM of the absorption components (given in C99)
and the electron temperature, provided by the optimal solution
(Table~3).  The FWHM is attributed to the effects of thermal Doppler
broadening and velocity gradients in the MHD flow.  The component due
to velocity gradients, $\Delta v$, is obtained by making a correction
to the FWHM by subtracting the thermal broadening which we take to be
$v_s=\sqrt{2kT_e/m_i}$, and by taking the ion mass $m_i$ to be the
average mass of carbon and nitrogen atoms. This average is taken
because the FWHMs listed in C99 were the result of averaging the FWHM
determined from the N{\sc~v} and C{\sc~iv} absorption lines. Table~3b
lists $v_{s}$ and $\Delta v$, defined by $\Delta v = FWHM-v_{s}$.  All
of the $\Delta v$ values are positive, therefore, some of the line
width cannot be attributed solely to thermal Doppler broadening.
Equation~7 is used to determine the spatial extent $\Delta R$ of each
component via
$$\Delta R = R\left(v_{obs}-{\Delta v\over 2}\right)-R\left(v_{obs}+{\Delta 
    v\over 2}\right) = $$
\begin{equation}
4R\left({\Delta v\over 2v_{obs}}\right)
      \left(1-\left[{\Delta v\over 2v_{obs}}\right]^2\right)^{-2}. 
\end{equation}
\begin{figure}[ht]
\vbox to2.6in{\rule{0pt}{2.6in}}
\includegraphics{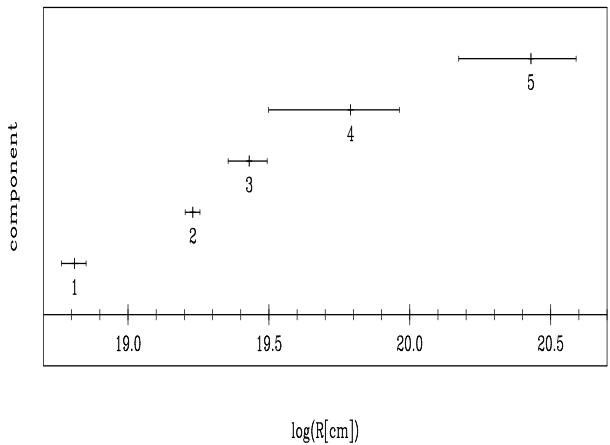} 
\caption{The location and width of the discrete
absorption components of the optimal model for warm absorbing gas in NGC~5548.}
\end{figure}

Numerical values for $\Delta R$ in each component are also listed in
Table~3b.  Figure~3 shows the width and location of each component for
the optimal solution plotted in $\log(R)$. Gaps between the components
indicate that they are well separated (except for possibly components 3
and 4), and that $\Delta R$ values grow monotonically with $R$.  For
the first three components $\Delta R/R<0.32$, but it is of order unity
for components 4 and 5.  According to the EBS model, the ionization
parameter will vary across the component by
$U_{front}/U_{back}=\sqrt{(R+\Delta R/2)/(R-\Delta R/2)}$.  The values
of $R$ and $\Delta R$ for each component (Table~3b) yield percentage
differences in $U$ measured at the front and back sides of each
component. These are $\sim 11\%$, $6\%$, $17\%$, $71\%$ and $61\%$ for
components 1 through 5, respectively.  Thus our approximation in
Equations~8, 9 and 10 is reasonable only for the first three
components.  For components 4 and 5 we estimate that the effect will
result in an error of about $-0.1$~dex in the N{\sc~v} and C{\sc~iv}
column densities, and as much as $+0.4$~dex uncertainty in the oxygen
column densities. Fortunately, these errors are minor in those
components where oxygen ions predominate (i.e., 1 and 3), and, therefore, 
the breakdown
of our assumption of constant $U$ across individual components does not
affect much the predicted integrated oxygen column densities.

An estimate for the volume filling factor is obtained from the column
density.  Using $n\propto R^{-1.5}$ as a density profile, the integral
in Equation~8 is given by $\int_{\Delta R}^{ }n dR =n(R)2R[(1-\Delta
R/2R)^{-1/2}-(1+\Delta R/2R)^{-1/2}]$, where $n(R)$ and $R$ are the
density and radius at the center of each component of width $\Delta R$.
Hence
$$N(X_{i})\approx $$
\begin{equation}
\epsilon f(X_{i})a_{X}n(R)2R[(1-\Delta R/2R)^{-1/2}
-(1+\Delta R/2R)^{-1/2}].
\end{equation}
If $n(X_{i})$ is the density of ion $X_{i}$ at $R$, then
\begin{equation} 
n(X_{i})=f(X_{i})a_{X}n(R).
\end{equation}
Substituting Equation~16 into Equation~15, letting $q=\Delta R/2R$ and 
solving for $\epsilon$ gives
\begin{equation} 
\epsilon \approx \frac{N(X_{i})}{n(X_{i})\Delta R}
\frac{q(1-q^{2})^{1/2}}{(1+q)^{1/2}-(1-q)^{1/2}}.
\end{equation}

The quantity $N(X_{i})/{n(X_{i})\Delta R}$ is the filling factor, when
the density is constant throughout the component.  The term involving
$q$ makes a correction for the variation in density across a
component.  The correction factors, however, are close to unity.  For
components 1 through 5, they are 0.994, 0.998, 0.984, 0.844 and 0.871,
respectively.  This requires the choice of an ion to define $\epsilon$,
and we choose C{\sc~iv} since it is already used as one of the solution
parameters.  We note that N{\sc~v} or even a blend of the C{\sc~iv} and
N{\sc~v} column densities could be used as well, but the differences in
calculated $\epsilon$ are not significant.

Calculated values of $\log(\epsilon)$ for the optimal solution range
from $-5.58$ to $-1.37$ (Table~3b). This means that an absorption
component is clumpy and has a filamentary structure.  In other words,
each of the five distinct absorption components in NGC~5548 consists of
filaments which move along similar trajectories and have probably a
common origin in the disk.  On the other hand, we presume that the
distribution of points of origin of any given absorption component in
the disk plane is random.  The clumpiness of individual components
combined with the geometry and non-radial nature of the MHD flow has an
important implication for the longevity of UV absorption structures in
AGNs, which we discuss in the next section.

\subsubsection{Temporal Changes in the UVX Absorbing Column}

Additional information about the character of the absorbing column of
gas is inferred from the temporal behavior (or lack thereof) of UV and
X-ray absorption.  Changes in absorption lines in principle can be due
to the motion of material into or out of the observer's line-of-sight,
or it may be due to changes in the ionization structure of the gas
brought about by changes in the incident continuum or both.  The X-ray
absorption line variability has been discussed in Reynolds \& Fabian
(1995), Reynolds (1997) and Otani et al.\ (1995), where \textit{ASCA}
observations of MCG~6-30-15 show apparent anti-correlation of the
O{\sc~viii} edge depth with an increase in the continuum level.
A similar conclusion was reached by Reynolds \& Fabian (1995).
Observations of UV absorption lines in AGNs also reveal that they
change with time.  Shull \& Sachs (1993) used {\it IUE} data to show
that the C{\sc~iv} absorption equivalent width in NGC~5548 responds to
continuum changes in an anti-correlated fashion on a time scale less
than about 4 days (this value is likely to be rather uncertain due to
the fact that the temporal sampling of this monitoring campaign was
four days).  UV observations of other AGNs spaced over many months reveal
that absorption lines change in depth with time, but the relationship
with the continuum is not clear.  While the depth of an absorption line
may vary, there are no unambiguous observations of a change in the
velocity centroid of the line (Weymann\ 1997), which argues against
purely radial motion, since one would expect clouds moving only
radially to accelerate along the line-of-sight, resulting in time with
a shift in the velocity centroid. This is in agreement with our
modeling of gas dynamics in the BLR of NGC~5548 (Paper~1) using the EBS
wind, which ruled out a purely radial motion as well. Here we focus on
the temporal phenomena associated with UV and X-ray absorption in
AGNs.

Our analysis of the discrete absorption components in the previous
section suggests that each component is an agglomeration of filaments
moving along the same trajectory. The heuristic model is that the wind
filaments are confined by the ambient magnetic field and are extended
along the field lines. The exact details of gas loading onto the
magnetic lines are subject to future work. Filaments which rotate out
of the line-of-sight are replace by their neighbors, so the velocity
centroid of the absorption component is not expected to change with
time.  The time scale $t_c$ over which we expect an absorption
structure to survive is given by $t_c=\Delta R/v_{\phi}$, where $\Delta
R$ is the width and $v_{\phi}$ is the rotational velocity of the
absorption component. For the first UV absorption component in
NGC~5548, where $t_c$ has the \textit{smallest} value, we find
$v_{\phi}\sim10^3~\rm{km~s^{-1}}$ and $\Delta R \sim 10^{19}\ {\rm
cm}$, leading to $t_{c}~\sim\/~3,000$ years.  The proposed model can,
therefore, explain the observed long constancy of the velocity
centroids of UV absorption structures.

Given the long dynamical time scale, observed temporal changes in the
absorbing column are attributed instead to changes in the
photoionization structure of the absorbing gas, rather than a change in
gas density along the line-of-sight, and are presumably due to
fluctuations in the incident continuum.  The response of the
absorption within the EBS wind to continuum fluctuations is
investigated by considering the effect of a sudden drop in the
continuum level on the column density of C{\sc~iv}, O{\sc~vii} and
O{\sc~viii}.  C{\sc~iv} is chosen to compare with observation (Shull \&
Sachs 1993), and O{\sc~vii} and O{\sc~viii} are chosen to compare
and contrast the response of our model of NGC~5548 with observations of
MCG~6-30-15.

In the above scenario, ions will begin to recombine with the drop in
the continuum intensity.  However, the time scale for
recombination as given by Equation~1 does not account simultaneously for  
the cascade into the population of $X_{i}$ ions from
the population of $X_{i+1}$ ions, and the cascade out of the population
of $X_{i}$ ions into the population of $X_{i-1}$ ions. Equation~1 is
hence misleading and is replaced for consistency by
\begin{equation} 
t(X_{i})=\frac{1}{\alpha(X_{i})n_{e}[\frac{f(X_{i+1})}{f(X_{i})}-
\frac{\alpha(X_{i-1})}{\alpha(X_{i})}]}.
\end{equation}
The value of $t(X_{i})$ in Eq.~18 establishes the \textit{minimum} time, that gas
will respond to a decrease in the continuum, and depends on the local
density {\it and} on the ion population.  As a single measure within a 
component, we take the average ion density-weighted recombination time which is
\begin{equation} 
<t(X_{i})>=\frac{\int_{\Delta R}^{ }t(X_{i})n_{X_{i}}\epsilon dR}{N_{X_{i}}}.
\end{equation}
This definition weights more heavily recombination times in regions
with higher column density contributions.  Assuming the validity of
Equations~$8-10$, $n_{X_{i}}$,$n_{X_{i+1}}$ and $n_{e}$ are all
proportional to $R^{-1.5}$.  In addition, $T_{e}$ {\em within} each
component is approximately independent of $R$ and, therefore,
$\alpha(X_{i})$ and $\alpha(X_{i-1})$ are also independent of $R$.
The result is that $t(X_{i})n_{X_{i}}$ is independent of $R$ and may be
factored out of the integral. This gives
\begin{equation} 
<t(X_{i})>=\frac{t(X_{i})\mid _{R}n_{X_{i}}(R)\epsilon\Delta R}{N_{X_{i}}},
\end{equation}
where $t(X_{i})\mid _{R}$ and $n_{X_{i}}(R)$ are $t(X_{i})$ and $n_{X_{i}}$,
respectively, evaluated at $R$.  Hence, $<t(X_{i})>$ may be positive or
negative, depending on whether or not the column density of ion $X_{i}$ is
anti-correlated or correlated with a decrease in the ionizing continuum,
respectively.  

Calculated values of $<t(X_{i})>$ for C{\sc~iv}, O{\sc~vii} and
O{\sc~viii} in NGC~5548 are listed in Table~4. Not surprisingly,
these recombination time scales differ from those calculated
from Equation~1 --- an issue sometimes ignored in the
literature. For a {\em fixed} electron density, the ionic
distributions as well as the effect of recombinations to lower
ionization states can produce a wide range of the recombination time
scales for the various metallic ions. In Table~4, the computed
value of $<t(X_{i})>$ for each ion differs from component to
component.  The recombination time scale for C{\sc~iv} spans a factor
of $\sim 5,000$.  The ranges for O{\sc~vii} and O{\sc~viii}, span
factors of $\sim 600$ and $\sim 50$, respectively.  Whether or not
changes of ion $X_{i}$ in a single component have a significant effect
on the total ionic column density over all components of ion $X_{i}$,
depends on the relative contribution of $X_{i}$ in the single
component. (All reported variations in ionic UVX column densities were
measured over the total value, due to insufficient spectral
resolution.)  For this reason the percent contributions of C{\sc~iv},
O{\sc~vii} and O{\sc~viii} are also shown in Table~4.  The values are
used to define the column density-weighted mean recombination time over
all components, $<<t(X_{i})>>$ (Table~4).  The individual recombination
times in each component serve as lower bound response time scales to
changes in the ionizing continuum flux, and $<<t(C{\sc~iv})>>$ may be
compared in a crude manner with observations.  In our model,
$<<t(C{\sc~iv})>> \approx +3.7$ days is reasonably consistent with the
results of Shull \& Sachs (1993), namely, that C{\sc~iv} absorption
equivalent width is anti-correlated with continuum variations with a
delay of roughly four days.

Long term X-ray observations of NGC~5548 do not yet exist, so the time scales
reported in Table~4 for O{\sc~vii} and O{\sc~viii} cannot be tested.  Instead
we compare the results of this model to the observations of MCG~6-30-15.  In
our model of the UVX absorption flow in NGC~5548, the bulk of the contribution
to the column density of these two oxygen ions resides in components 1 and 3.
However, the values of $<<t(O{\sc~vii})>>$ and $<<t(O{\sc~viii})>>$ are
considerably skewed by the extremely long recombination time scale of component
5.  As a conservative measure, we, therefore, consider only components 1 and 3.
In these components, the column density contributions are roughly equal for
both O{\sc~vii} and O{\sc~viii}, but the recombination times in the first
component are over 10 times shorter.  These recombination times, which
represent the minimum possible response time scale for a continuum drop, are
$<t(O{\sc~vii})>\approx -3$ days and $<t(O{\sc~viii})>\approx -50$ days.  (The
O{\sc~vii} and O{\sc~viii} edge optical depths are correlated with the ionizing
continuum intensity.)  We contrast this with observations of MCG~6-30-15 (see
section~2).  During the second half of the $\sim4$ day observation period of
this object, the X-ray continuum dropped by a factor of 2.  The response was
that the O{\sc~viii} optical depth anti-correlated with the continuum on a time
scale of $\sim 10^{4}$~s, but there was no observed change in the O{\sc~vii}
optical depth.

A possible explanation for the observed behavior of the O{\sc~viii}
edge optical depth in MCG~6-30-15 is that the absorbing gas in
MCG~6-30-15 is more highly ionized than in NGC~5548.  The observed
continuum in MCG~6-30-15 is much harder than in NGC~5548, and the WA
gas may, therefore, support a large population of O{\sc~ix} ions.
Equation~18 allows then for a complex dependency of the recombination
time scales, which might explain the differences between the
predicted behavor of NGC~5548 and the observed behavior of
MCG~6-30-15.  Long looks by {\em Chandra} and other advanced X-ray
telescopes should establish the relationships between the continuum and
WA optical depths in AGNs.  

\section{A GENERIC MODEL FOR WARM ABSORBING GAS}

The discrete UVX absorber model for NGC~5548 does not allow for its
direct application to other objects. This is mainly because detailed
kinematic data is not available, the model does not provide for the UVX
absorption dependence on the aspect angle, and because it does not
specify the appropriate boundary conditions {\it not} along the
line-of-sight.  To generalize the model, we proceed by spatially
smoothing the contributions from individual UVX kinematic components of
our model and by calculating the optical properties of the MHD flow
along a number of aspect angles.  The resulting generic model is used
to estimate orientation effects on the soft X-ray absorption properties
of AGNs. The general aspects of thermal stability of the WA gas in this
model are discussed in the Appendix.

\subsection{Column Densities and physical conditions}

\begin{figure}[ht]
\vbox to4.0in{\rule{0pt}{4.0in}}
\includegraphics{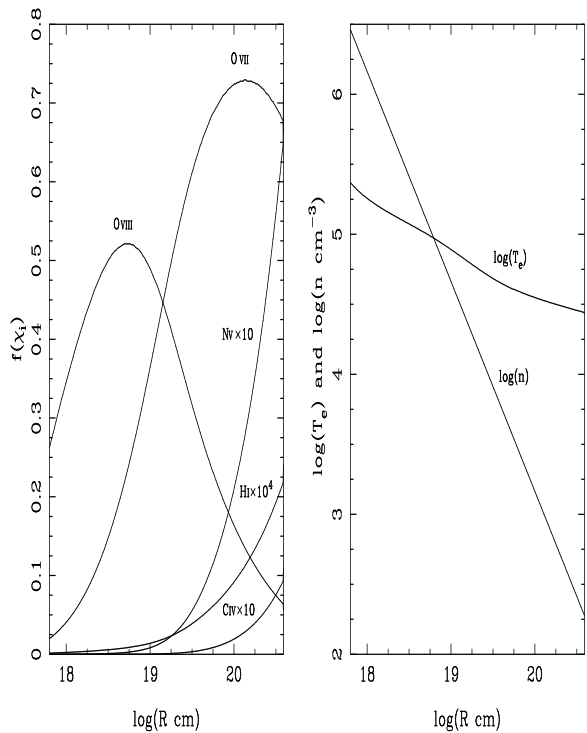}
\caption{{\it Left:} Plot of the ion fraction as a function of radius
for the continuously distributed model of NGC~5548. For clarity,
C{\sc~iv} and N{\sc~iv} values have been multiplied by a factor of
10.  H{\sc~i} has been multiplied by a factor of
$10^4$. {\it Right:}  Logarithmic plot of the particle density 
($\rm{cm^{-3}}$) and temperature ($^\circ$K) as a function of radius (cm) for
the continuously distributed model of NGC~5548.}
\end{figure}

We first build a quasi-continuous model of MHD flow in NGC~5548 by
smoothing the kinematic components along the line-of-sight, by imposing
observational constraints on the total column density of hydrogen, and
by retaining the $\propto R^{-3/2}$ density profile from the discrete
component model.  For simplicity, the outer edge of the continuous flow
was set at ${\rm log}R\approx 20.6$ which coincides with the outer edge
of the fifth component.  Two free parameters, i.e., the volume filling
factor and the inner radius, which we used as Cloudy input parameters,
were tuned to fit observations of the column densities of
$N(C{\sc~iv})$, $N(N{\sc~v})$, $N_{H}$, $N(O{\sc~vii})$ and $N(O{\sc
viii})$.  A reasonable fit to the ionic column densities is found for
$\log(R_{1})\approx 17.7$ and $\epsilon\sim 10^{-3}$ (Table~2). The
first plot in Figure~4 clearly illustrates ionization stratification,
with the ion fraction O{\sc~viii} achieving a maximum at smaller radii,
O{\sc~vii} at intermediate radii, and both C{\sc~iv} and N{\sc~v}
peaking near the outer edge.  The second plot in Figure~4 shows the
density and electron temperature as a function of distance.  Note the
temperatures range from $\sim 1.6 \times 10^{4}\rm K$ at larger radii
to $\sim 2 \times 10^{5} \rm K$ at smaller radii, where the bulk of the
WA resides.

\subsection{Orientation Effects in AGNs and Soft X-Ray Absorption}

\begin{figure}[ht!]
\vbox to4.6in{\rule{0pt}{4.6in}}
\includegraphics{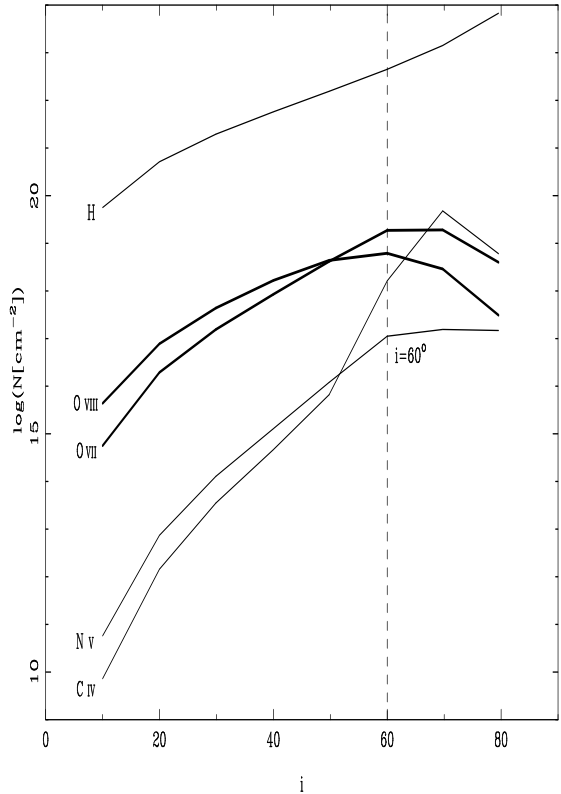}
\caption{Plot of the log of the column densities ($\rm{cm^{-2}}$)
of N{\sc~v}, C{\sc~iv}, H, O{\sc~vii} and O{\sc~viii} as a function of
observer aspect angle $i$ for the continuously distributed model. 
Values with angles greater than $60^{o}$ do not
include contributions to the column from optically thick BLR clouds.
The H column is therefore most likely to be an order of magnitude (or
more) higher when $i>60^{o}$.}
\end{figure}

We now investigate the orientation effects in the continuous wind model
of NGC~5548 by varying the viewer aspect angle.  Figure~5 shows the
column densities of $N(N{\sc~v})$, $N(C{\sc~iv})$, $N(H)$,
$N(O{\sc~vii})$ and $N(O{\sc~viii})$ as a function of viewer aspect
angle $i$.  The density profile along each aspect angle was adjusted so
that the MHD flow is normalized to the solution for NGC~5548 (Table~2),
which corresponds to the aspect angle $i=40^\circ$.  Absorbing column
densities seen by an observer with aspect angle $i>60^\circ$ are within
the angular wedge shadowed by the BLR, i.e., by hydrogen column
densities in excess of $10^{23}\ {\rm cm^{-2}}$.  Although we restrict
our discussion to angles at $60^\circ$ or below, and to smaller column
densities compatible with those in NGC~5548, larger column densities
cannot be excluded for other objects. We adopt the generic model for
NGC~5548 as a template, when applied to other Seyfert~1 galaxies.

Figure~5 shows that $N(O{\sc~vii})$ and $N(O{\sc~viii})$ decrease with
decreasing $i$, making detection of the oxygen edges more difficult for
small $i$.  Using the \textit{instrument} error bar in R97, and
assuming a $2\sigma$ detection criteria, absorption edges of O{\sc~vii}
and O{\sc~viii} will be lost for aspect angles less than $i\sim
30^\circ$ with the disk axis.  Since no objects will be seen for
$i>60^\circ$, due to obscuration, we restrict our analysis to angles
$i<60^\circ$.  In this case, if all Seyferts are taken as randomly
oriented templates of NGC~5548, then about $\sim 73\%$ of them will
have detectable WA gas.

Figure~5 also shows that $N(O{\sc~vii})$ drops faster than
$N(O{\sc~viii})$ when viewed at smaller aspect angles, and drops below
$N(O{\sc~viii})$ for angles less than $50^\circ$.  The fraction of
Seyfert~1 galaxies with detectable WA showing $N(O{\sc~viii}) >
N(O{\sc~vii})$ can be estimated from the ratio of the solid angle
subtended on a sphere between the detection limit $30^\circ$ and
$50^\circ$, and the solid angle between $30^\circ$ and $60^\circ$.
This ratio is about $61\%$, for current instrument sensitivity.  For
comparison, 8 out of 16 Seyfert~1 galaxies listed in R97 (50\%) have
$N(O{\sc~viii}) > N(O{\sc~vii})$. (We used only Seyfert~1s from R97
that have definite error bounds on both the tabulated values of optical
depth $\tau_{O{\sc~vii}}$ and $\tau_{O{\sc~viii}}$.)  Finally, we note
that the predicted range of $N_{H}$ varies from $\sim 10^{21}\ {\rm
cm^{-2}}$ to $\sim 10^{23}\ {\rm cm^{-2}}$ within the observable range
of aspect angles, and this compares favorably with the inferred range
of $N_{H}$ in R97.  Of course, one can question this approach, when one
object is used as a template to describe the whole population of
Seyferts, and the Seyfert sample in R97 is small.  The model's
consistency with the observations in these respects is nevertheless
interesting. Taken at face value, the ratio of optical depths of
O{\sc~vii} and O{\sc~viii} ions can serve as a new diagnostic for AGN
aspect angles.

\subsection{Diffuse Emission from the UVX Absorber}

The UVX absorbing gas reprocesses and scatters the incident continuum
into diffuse emission across the UV and soft X-rays.  We used our
``generic'' UVX absorber for NGC~5548 to estimate the magnitude of this
emission.  We computed Cloudy models for thermal local line widths and
one, in which microturbulence broadened the local line width
($\sigma_{turb} = 100\ {\rm km\ s^{-1}}$).  In the latter scenario,
optically thick lines are desaturated, and the cross-section for photon
pumping of the line's upper level (continuum resonance line scattering)
is elevated, enhancing the line intensities.  Photon pumping may be
especially important for the soft X-ray lines because of their small
Boltzmann factors in photoionized gas.  However, even under the
assumption that the observer can see the gas represented by our
generic UVX absorber distributed over $4\pi$ steradians over the
continuum source, the diffuse emission is relatively small.  This is
expected, given the reported small optical depths of the absorber in
this object.  We find that this gas contributes at most $\sim$~20\% to
the observed UV {\em narrow} line fluxes of Ly$\alpha$, C{\sc~iv}, and
N{\sc~v} in NGC~5548 (Goad \& Koratkar 1998). Emission of this
significance may be sufficient to partially fill in some of the
absorption troughs.  Depending upon the covering fraction and the
contribution from photon pumping the diffuse soft X-ray line emission
could be also be significant in comparison to the absorption edge
depths (Netzer 1993).  We will have to await the results from the
upcoming missions of {\em Chandra}, {\em XMM}, and {\em
Constellation-X} to get hard observational constraints from this
portion of the spectrum.

\section{SUMMARY}

We have applied the EBS hydromagnetic wind model from a clumpy
molecular accretion disk to the problem of UVX-ray absorption systems
observed in Seyfert galaxies. We have studied the state of the UVX
absorbing gas, with a particular emphasis on NGC~5548, where five
discrete UV absorption systems seen in this object were modeled.
Extending our previous work on the dynamics of the BLR gas in NGC~5548
(Paper~1), we first inferred the location of each component relative to
the continuum source.  By estimating the ratio of the ion fractions of
N{\sc~v} to C{\sc~iv} from the observed column densities and the use of
photoionization modeling, we have determined possible values for the
density, electron temperature and the ion fractions of C{\sc~iv},
N{\sc~v}, O{\sc~vi}, O{\sc~vii} O{\sc~viii}, H and H{\sc~i}.
Optimization over the set of possible values yielded the ion column
densities in each kinematic component and allowed us to test the radial
and polar density distributions of the WA in NGC~5548, within the EBS
framework. We have also estimated the size and volume filling factor of
each component, using the FWHMs and electron temperatures, and found
that the flow in each component is clumpy. X-ray absorption by
O{\sc~vii} and O{\sc~viii} column densities seen in NGC~5548 have been
accounted for in the five UV absorption components, though main
contribution to the X-ray absorption comes from two components. We,
therefore, were able to explain the UVX absorption columns within the
framework of the same dynamical model. An additional important point is
that the model parameters used here are the best fitting parameters
from Paper~1 to explain the broad emission-line variability in NGC~5548
during the 1989 and 1993 observing campaigns.

Secondly, we find that the WA gas in NGC~5548 lies at larger radii from the
central source than the BLR gas and at larger altitudes above the disk.  This
means that WA exists in the outer parts of the hydromagnetic disk wind, and as
such is {\it not} a continuation of the BLR flow. As a result we do not 
expect it to contribute to absorption
along the line of sight unless the aspect angle is near the obscuring torus.
With regard to the spatial extent of the WA gas we find it extends in both
radial and polar directions from the central continuum source, and is ionization
stratified.

Thirdly, we
have modeled the WA in NGC~5548 also as a continuous flow, in order to
investigate the generic properties of the UVX absorption outflows in
Seyfert galaxies. We find that orientation effects may fully account
for the fraction of objects with detectable WAs. Furthermore, the model
predicts that the ratio of optical depths of O{\sc~viii} to O{\sc~vii}
can serve as a new diagnostic of orientation of an AGN, namely of its
rotation axis --- an issue of a particular interest in AGN theory. In
addition, we find that the UV line emission from the UVX flow probably
accounts for no more than $\sim 20\%$ of the NLR emission in lines,
such as Ly$\alpha$, C{\sc~iv}, N{\sc~v}, and O{\sc~vi}, though diffuse
emission in the soft X-rays may be significant under some conditions.

A thermal stability analysis has been carried out for the model.  We
find that all of the five components in NGC~5548 are stable, though two
components (1 and 3) lie near an unstable, to isobaric perturbations,
region on the S-curve. By considering
the effects of magnetic field on the thermal stability of the WA, we
find that these components can be further thermally stabilized by a
modest field, even in the presence of substantial continuum
fluctuations, even when considering extremes in SED or metallicity.
Such stabilization by magnetic fields may explain the ubiquity of WA
gas in AGNs.

As more objects are observed with more sensitive instruments (e.g.,
{\em Chandra} and {\em XMM} coupled with the {\em HST}), better
statistical tests of the wind model can be carried out, involving
determination of the fraction of AGNs with detectable WA gas, the
frequency of objects in which the O{\sc~vii} column density is larger
than the O{\sc~viii} column density, and a determination of the 
range of $N_H$.  Future higher resolution X-ray spectra
(\textit{ASCA}, \textit{XMM}) will provide additional kinematic
information, such as oxygen edge and resonance line absorption
velocities and help to verify the link between X-ray and UV
absorption.

\acknowledgments
We gratefully acknowledge illuminating discussions with Roger
Blandford, Mike Crenshaw, Arieh K\"onigl, Richard Mushotzky and Chris
Reynolds.  We thank Gary Ferland for the use of Cloudy.  This work was
supported in part by NASA grants NAG5-3841, WKU-522762-98-06,
WKU-521782-99-04 and HST AR-07982.01-96A.

\appendix
\section{THERMAL STABILITY OF WARM ABSORBING GAS}

The WA gas is found in a wide range of densities, temperatures,
ionization and metallicity. It seems plausible, that this ubiquity of
WAs in AGNs is related to their thermal stability.  Permeating magnetic
fields which are expected to play an important dynamical role in AGNs,
can affect the thermal stability of the absorbing gas. Therefore, along
with Field (1965) and EBS, we discuss the general aspects of thermal
stability of magnetized gas and its application to NGC~5548 and to a
generic model of the WA, subject to AGN radiation field.

\subsection{The ``S-curve'' and Thermal Stability: General Considerations}

\begin{figure}[ht!!!!!]
\vbox to4.1in{\rule{0pt}{4.1in}}
\includegraphics{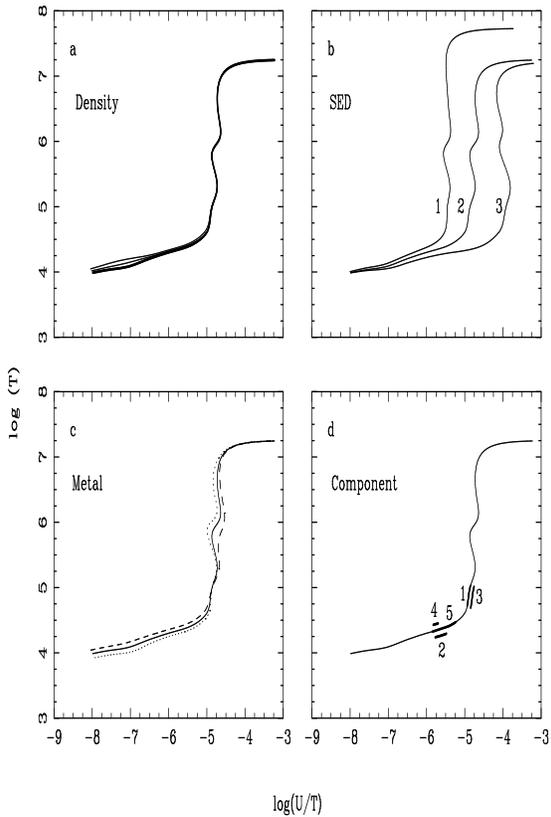}
\caption{S-curves for various parameters. {\it (a)} S-curves for the 
SED of NGC~5548 for densities $10^{5}$,
$10^{6}$, $10^{7}$ and $10^{8}$ $\rm{cm^{-3}}$.  The thick line
contains both the $10^{5}$ $\rm{cm^{-3}}$ and $10^{6}$ $\rm{cm^{-3}}$
S-curves.  Note that
above $10^{5}$ K all the S-curves merge, illustrating that the
S-curve is relatively insensitive to differences in density. {\it (b)}
S-curves corresponding to different incident
continuum SEDs.  The S-curve labeled 1 is for the SED of
MCG~6-30-15; the S-curve labeled 2 is for the SED of NGC~5548; the
S-curve labeled 3 is for the Mathews \& Ferland (1987) composite
spectrum of quasars.  The hardness of the SED decreases from left to
right. {\it (c)} S-curves for the SED of
NGC~5548 with three different metal abundances.  The solid line
corresponds to solar abundances; the dashed line is for a metal
abundance of $0.5\times$solar; the dotted line is for an abundance of
$2.0\times$solar. {\it (d)} Location of the 5
discrete absorption components observed by C99 on the S-curve for
NGC~5548.  Segments are offset for clarity.  Components 1 and 3 lie
close to an unstable (to isobaric perturbations) portion of the
S-curve.}
\end{figure} 

For a gas in thermal equilibrium, the heating rate per unit volume,
$G(n,T)$, is balanced by the cooling rate per unit volume,
$\Lambda(n,T)$. If $H$ is the net heating rate defined by
\begin{equation}
H \equiv G(n,T)-\Lambda(n,T),
\end{equation}
then equilibrium occurs when $H=0$.  If gas is slightly perturbed,
i.e., $H\neq0$, the gas will either heat up or cool down. The change in
the gas temperature defines whether it is thermally stable (Field 1965;
Krolik, McKee \& Tarter 1981).  The equation $H=0$ is more conveniently
treated when mapped to the ${\rm log}~T - {\rm log}~U/T$ plane where
the equilibrium curve has roughly an S-shape.  Thermal stability of gas
depends on the details of the S-curve, {\it and} the path that a
perturbation is allowed to move along.  Figure~6a displays a series of
S-curves corresponding to different hydrogen gas densities for the SED
of NGC~5548. The curves are relatively insensitive to the gas densities
considered and become virtually indistinguishable above $T\approx
10^5$~K.

\begin{figure}[ht!]
\vbox to4.1in{\rule{0pt}{4.1in}}
\includegraphics{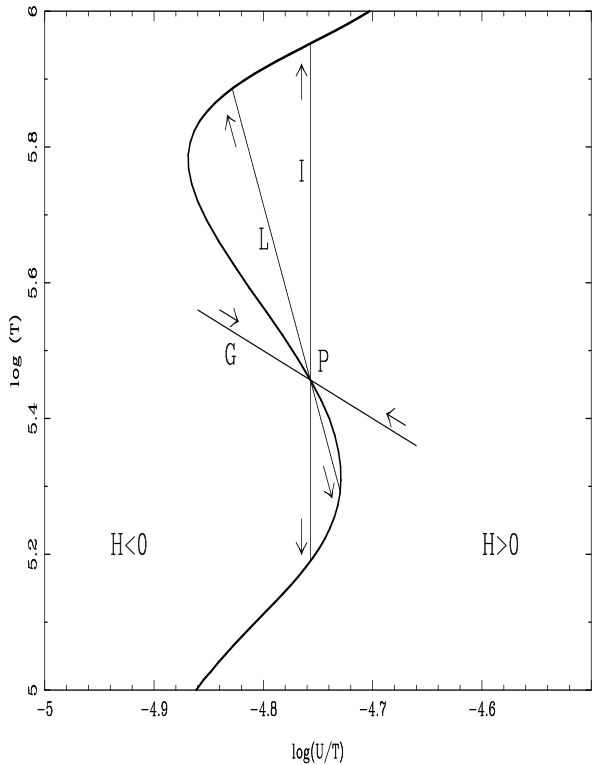}
\caption{Schematic depiction of an unstable (to isobaric
perturbations) portion of the S-curve.  To the left of the S-curve
cooling exceeds heating.  To the right of the S-curve heating exceeds
cooling.
Different perturbation constraints (paths G, L, and I) are shown
crossing the point P.  Arrows show the direction gas will evolve along
a constraint when perturbed off the point P. Path I corresponds to a
thermally unstable
isobaric perturbation at P. Path L has a slope
less than the S-curve at point P, therefore, perturbations constrained on
this path are unstable.  Path G has a slope greater than the S-curve at
P, so perturbations constrained to this path are stable.  }
\end{figure}

To analyze thermal stability of gas, one must be able to follow the
path of the perturbed gas relative to the S-curve. Figure~7 shows a
small segment of the S-curve for NGC~5548, having a negative slope in
the vicinity of point P.  To the right of the S-curve, the net heating
$H$ is positive, i.e., perturbed gas always heats up.  To the left of
the S-curve, the net heating $H$ is negative, meaning that perturbed
gas always cools down.  Crossing the S-curve at the equilibrium point P
are three paths (G, L and I), representing different alternatives for
an evolving perturbation off the point P. Path~I, a vertical line,
represents an isobaric perturbation\footnote{If luminosity and radius
are fixed then $U/T \propto 1/nT$, therefore, $U/T=constant$ implies
$nT$ is constant.  Hence, the gas moving on vertical lines in the $T$
vs.~$U/T$ plane evolves isobarically.}. G and L represent perturbations
constrained to paths with slopes greater (less negative) and with slope
less (more negative) than the slope of the S-curve at P, respectively.
Along I and L, a small perturbation off the S-curve leads to a dramatic
change in temperature.  Perturbations constrained to move along G, on
the other hand, are thermally stable since they are driven back to P.
Hence, the thermal stability of the gas depends on the slope of
developing perturbation in ${\rm log}~T - {\rm log}~U/T$ plane. The gas
is thermally stable when this slope is greater (less negative) than the
slope of the S-curve at P.  For an ionized gas permeated by a magnetic
field, as shown below, perturbations move along inclined paths in this
diagram, e.g., as G or L in Figure~7.  Since stability depends on the
relative steepness of the S-curve, as compared to that of the the
perturbation path, we first investigate the factors which influence the
shape and, therefore, the slope of the S-curve.

\subsection{Factors Affecting the S-curve: SED and Metallicity}

The overall shape of the S-curve is affected by the SED.  Figure~6b,
shows the S-curves for three different SEDs with labels corresponding
to SED with photon index $\Gamma =1.8$ and energy cutoffs between
13.6~eV and 40 keV (this SED was utilized in Reynolds 1995), to the SED
used in our modeling of NGC~5548, and to the typical quasar SED deduced
by Matthews \& Ferland (1987).  The sequence is in terms of decreasing
hardness of the continuum and the greatest effect is on the upper
branch of the S-curve. Here the temperature is mainly controlled by
Compton heating and cooling.  In the limit of large $\log(U/T)$, the
S-curve is asymptotic to the Compton temperature $T_C$ which is given
by $T_{C}=h<\nu>/4k$, where $<\nu>$ is the flux-averaged photon
frequency.  Higher $<\nu>$ corresponds to harder SED, and, therefore,
higher $T_C$, explaining the sequence of decreasing upper branch
temperatures.

The horizontal separation between S-curves (Fig.~6b) is due to decrease
in the continuum hardness, when the percentage of high energy photons
($h\nu >1$~Ryd) that heat the gas per unit frequency interval
decreases. Decreasing the hardness, allows the gas to cool down, unless
compensated for by increasing the number of high energy photons, which
is the same as increasing the ionization parameter.  Thus, for a {\it
given} value of $T$, the S-curve lies further to the right (i.e., it
has higher $U/T$ values which correspond to higher values of $U$) as
the hardness decreases. In spite of a wide range of hardness, the
slopes of the S-curves vary little, particularly in regions where the
slope is negative.  We further note that in all cases considered here,
the S-curve slopes is always {\it less than} $-1$, in regions where the
slope is negative.  This is an important point because, as we show
below, the path that gas perturbation treaded by a magnetic field
follows, has slope close to $-1$, resulting in thermal stabilization.

The bumps and wiggles in an S-curve are due to the metallicity of the
gas.  If AGNs are fueled with galactic ISM, its composition may reflect
intense localized star formation which is typically concentrated within
nuclear and circumnuclear regions and, therefore, uniformity of
composition is not guaranteed.  In addition, when grains are directly
exposed to an intense central continuum, their mantles will be destroyed and
only the graphite cores will be able to survive photo-desorption (Draine \&
Salpeter 1979) at distances and in the radiation field relevant for WA in
NGC~5548. This introduces the possibility of metallicity gradients within the
WA environment. Gas abundances are important, because at WA temperatures line
radiation by heavier elements, such as oxygen, constitute the dominant
coolant.  Thus metallicity can influence the thermal structure and stability
of the WA gas. This is demonstrated below.

Figure~6c shows S-curves for the SED of NGC~5548 for gas with three
different metallicities.  At temperatures above $T\approx10^5 K$,
higher metallicity curves are shifted further to the left than lower
metallicity curves in the ${\rm log}~T$ vs.  ${\rm log}~U/T$ plot.
This is because the presence of more metal ions gives the gas a net
higher energy absorption cross section than that at a lower
metallicity, so that a particular temperature is achieved with a lower
ionization parameter.  In addition, extra electrons provided by metal
enhancement are able to recombine with $H^{+}$ (and $He^{++}$), thus
the effective ionization of the gas is lower for a given ionization
parameter. For example, along a line of ${\rm log}~T\approx 5.85$, the
ionization parameter increases from roughly 0.85 to 1.14 to 1.25, as
the metallicity drops.  Below $T\approx 10^5$~K, the order is reversed
because the metal ions recombine and become more efficient coolers via
line emission than heaters via absorption.  Inspection of the Figure~6c
reveals that the slopes of the S-curves for different metallicities are
similar, and in the regions where the slopes are negative, they all are
less than $-1$.  So, metallicity differences, even as large as a factor
of 2 greater or less than solar, will not affect our conclusions about
thermal stability of magnetically-dominated gas.

\subsection{Thermal Stability of Magnetized Gas}

For isobaric perturbations, there are no dynamical effects and the
resulting evolution of the perturbation is controlled by thermal
effects only.  In the absence of a magnetic field, an isobaric path
corresponds to a vertical line (path I in Figure~7).  For this kind of
perturbation, regions where the slope of the S-curve is negative are
thermally unstable.  The location of the discrete UV/X-ray WA model is
shown in the ${\rm log}~T$ vs.  ${\rm log}~U/T$ plane (Fig.~6d).  The
temperature range of absorption components 1 and 3 lies close to a
region which is thermally unstable to isobaric perturbations, so
changes in continuum luminosity may drive these components into an unstable
region.\footnote{The temperature uncertainty for each component is
estimated by searching for the largest and smallest values temperature
values in the $(n,N(C{\sc~iv}))$ solution grid that are adjacent to the
optimal solution.}  Components 2, 4 and 5 are in a region where the
S-curve slope is positive, so these components are thermally stable.

The mere existence of kinematic components 1 and 3 in NGC~5548 hints
that the gas is thermally stable, despite being close to unstable
region. Such a stabilization mechanism can be provided by magnetic
fields permeating the gas (Field 1965; EBS). In the presence of
magnetic field, isobaric perturbations are constrained to evolve along
the curve defined by $nkT + B^2/8\pi = A$ where $A$ is a constant.  To
analyze thermal stability, we rewrite this expression in terms of $T$
and $\frac{U}{T}$ and show that when plotted in the ${\rm log}~T$ vs.
${\rm log}~U/T$ plane, the slope is approximately equal to $-1$.
Assuming magnetic flux freezing, the magnetic field scales with the gas
density $B^2=\phi^2n^2$, where $\phi$ is a constant.  Substitution into
the expression for total pressure gives $nkT + \phi^2n^2/8\pi = A$.  We
eliminate $n$ by solving
\begin{equation} 
\frac{U}{T} = \frac{L_{ion}}{<E_{ion}>4\pi R^2cnT}
\end{equation}
for $n$. This gives
\begin{equation}
n=\frac{E}{(\frac{U}{T})T},          
\end{equation}
where $E = L_{ion}/<E_{ion}>4\pi R^2c$. Substitution into the 
expression for total pressure gives
\begin{equation}
\frac{Ek}{\frac{U}{T}} + \frac{\phi^2 E^2}{8\pi T^2(\frac{U}{T})^2} = A,  
\end{equation}
and solving for $T$ yields
\begin{equation}
T = \frac{\frac{\phi\/E}{\sqrt{8\pi}}}{\sqrt{A(\frac{U}{T})^2-Ek(\frac{U}{T})}}.
\end{equation}
To thermally stabilize the gas requires
\begin{equation}
\frac{d~\log(T)}{d~\log(U/T)}>S,
\end{equation}
where $S$, the slope of the S-curve in the ${\rm log}~T$ vs.  ${\rm
log}~U/T$ plane, and the derivative on the left hand side of
Equation~A6 are evaluated at the point where Equation~A5 crosses the
S-curve, yielding
\begin{equation}
\frac{d {\rm log}~T}{d {\rm log}~(\frac{U}{T})} = -1-\frac{1}{2}\beta,
\end{equation}
where $\beta$ is the ratio of gas to magnetic field pressures.  Note,
that when the magnetic field $B\rightarrow 0$, then $\beta\rightarrow
\infty$, and $d {\rm log}~T/d {\rm log}~(\frac{U}{T})\rightarrow
-\infty$, recovering the strictly isobaric case.  At the other extreme
lies the limiting case of the cold EBS flow, in which the magnetic
pressure completely dominates the gas pressure ($\beta <<1$), so $d
{\rm log}~T/d {\rm log}~(\frac{U}{T}) \approx -1$.

\begin{figure}[ht]
\vbox to4.1in{\rule{0pt}{4.1in}}
\includegraphics{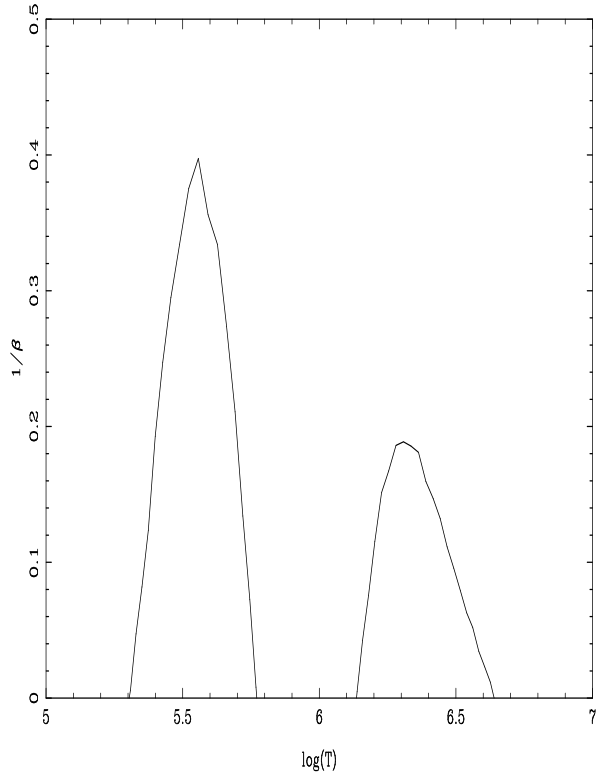}
\caption{$1/\beta$ vs.~$T$ for the S-curve of NGC~5548.
$\beta$ is the minimum value required to stabilize gas at the
corresponding temperature.}
\end{figure} 

The minimum value of $B$ required to stabilize the gas is found by
equating the slope of the S-curve to Equation~A7, and solving for
$\beta_{min}$.  Figure~8 shows $1/\beta_{min}$ values as a function of
temperature for the two regions of instability, for the SED of
NGC~5548.  For stable regions, no magnetic field is needed, so
$1/\beta_{min}\rightarrow0$.  Components 1 and 3 are close to the
thermally unstable temperature corresponding to the left peak in
Figure~8.  The maximum $1/\beta_{min}$ occurs at ${\rm log}~T=5.55$,
where $1/\beta_{min}=0.395$, giving
\begin{equation}
\frac{1}{\beta} = \frac{\frac{B^2}{8\pi}}{nkT} = 0.395,
\end{equation}
where $B$ and $n$ are the values for ${\rm log}~T=5.55$.  Thus,
according to Equation~A8, if the value of the magnetic pressure is
about $40\%$ of the gas pressure or larger the gas is stabilized.
Cloudy calculations and our wind model yield a density of ${\rm
log}~n=4.99$ at the center of component 1.  With this density a lower
bound on the magnetic field required for stabilization can be
estimated.  Solving Equation~A8 for $B$ gives
\begin{equation}   
B=\sqrt{8\pi \times 0.395 nkT}. 
\end{equation}
In the regime of the left peak, the gas is stable at all possible
temperatures, if $B \gtrsim 7.0\times 10^{-3}$~Gauss. It should be
noted, however, that the bound on $B$ is determined only from the
stability requirement. A more stringent requirement from the EBS model
is that the flow is a ``cold'' MHD flow, which means $1/\beta >>1$,
superseding the above criteria.  The implication is that the slope
given by Equation~A7 is nearly equal to $-1$.  We have noted above that
variations in SED or metallicity do not significantly change the slopes
of the negative parts of the S-curve, and that the slopes are always
less than $-1$, corresponding to case G of Figure~7, in which the slope
of the perturbation is larger than the slope of the S-curve. Thus, in
the case of NGC~5548, as described by the EBS flow, regions of the
S-curve that correspond to thermally unstable regions to isobaric
perturbations are thermally stabilized by the presence of a magnetic
field.

%

%
%
%

%
\newpage

\onecolumn
\centerline{\bf Table 1a}
\centerline{\bf \textit{ASCA} R97 X-ray Observational Summary NGC~5548}
\centerline{\bf (Adopted Target Values$^{a}$)}
\vspace{-0.2in}
\begin{center}
\begin{tabular}{ll}
\hline \hline 

$\log(N_{H})$            & $21.71$\\

$\log(N_{O{\sc~vii}})$  & $17.95$\\

$\log(N_{O{\sc~viii}})$ & $18.20$\\

\hline
\end{tabular}
\end{center}
\bigskip

\centerline{\bf Table 1b}
\centerline{\bf \textit{HST} C99 UV Observational Summary NGC~5548}
\centerline{\bf (Adopted Target Values$^{a}$)}
\vspace{-0.20in}
\begin{center}
\begin{tabular}{lllll}
\hline\hline
comp. & $\log[N(N{\sc~v})]$ & $\log[N(C{\sc~iv})]$ 
& $\log[N(H{\sc~i})]$ & $\log[\frac{f(N{\sc~v})}{f(C{\sc~iv})}]$\\

\hline 

1 & $14.30<$ & $13.04<$ & $14.13^{b}$ & $2.14>$ \\
 
2 & $13.78$ & $13.45$ & $13.88$ &  $0.91$  \\

3 & $14.59$ & $13.68$ & $14.25$ &  $1.49$  \\

4 & $14.81$ & $14.46$ & $14.52$ &  $0.93$   \\

5 & $14.04$ & $13.61$ & $13.11$ &  $1.01$   \\

\cline{1-4}

$\Sigma$ & $15.17$ & $14.62$ & $14.87$ \\

\cline{1-4}

\end{tabular}
\end{center}
 
$^{a}$ All values have assigned error bars of $\pm \log(2.0)$ except for
those with a $<$, which indicates a lower bound or $>$, which indicates an
upper bound.

$^{b}$STIS observation (Crenshaw \& Kraemer 1999)

\newpage

\centerline{\bf Table 2}
\centerline{\bf Model Solution for the Continuously Distributed}
\centerline{\bf Warm Absorber in NGC~5548}

\medskip

\begin{center}

\begin{tabular}{ll}

\hline\hline

parameter & log(value)$^a$ \\

\hline 

$\rm R_{min}$ & 17.75  \\

$\rm R_{max}$ & 20.59  \\

$\rm n_{min}$ & 6.455  \\

$\rm n_{max}$ & 2.253  \\

N(H) & 21.76    \\

N(H{\sc~i}) & 15.94\\

N(C{\sc~iv}) & 14.70 \\

N(N{\sc~v}) & 15.13 \\

N(O{\sc~vii}) & 17.94 \\

N(O{\sc~viii}) & 18.22 \\

$\epsilon$ & -2.716 \\

\hline

\end{tabular}
 
\end{center}

\centerline{$^a$ values in cgs units.}

\newpage

\medskip

\normalsize{
\centerline{\bf Table 3a}
\centerline{\bf Optimal EBS Solution Column Densities$^{a}$}
\vspace{-0.125in}
\begin{center}
\begin{tabular}{cccccccc}
\hline\hline
 comp. & C{\sc~iv} & N{\sc~v} & O{\sc~vi}  
 & O{\sc~vii} & O{\sc~viii} & H
 & H{\sc~i}\\
 
\hline

1& $13.50$ & $14.30$ & $16.15$ & $17.73$ & $17.84$ & $21.29$ & $15.57$\\

2& $13.45$ & $13.48$ & $14.57$ & $14.51$ & $13.00$ & $18.14$ & $14.11$\\

3& $13.62$ & $14.41$ & $16.25$ & $17.80$ & $17.88$ & $21.33$ & $15.66$\\

4& $14.46$ & $14.51$ & $15.65$ & $15.65$ & $14.20$ & $19.22$ & $15.14$\\

5& $13.49$ & $13.74$ & $15.11$ & $15.56$ & $14.55$ & $18.88$ & $14.35$\\

\hline

$\Sigma$ & $14.62$ & $14.94$ & $16.58$ & $18.07$ & $18.16$ & $21.61$ & $16.00$\\

\hline
\end{tabular}
\end{center}
\vspace{-0.2in}
\hspace{1.40in} $^{a}$Logarithims of column densities in units of $\rm cm^{-2}$

\vspace{1.0in}

\centerline{\bf Table 3b}
\centerline{\bf Optimal EBS Solution Physical Parameters$^{a}$}
\vspace{-0.125in}
\begin{center}
\begin{tabular}{cccccccccccc}
\hline\hline   
comp. & $\rm v_{obs}^{b}$  & FWHM$^{b}$
& $\rm v_{s}$
& $\Delta \rm v$
&  $\frac{f(N{\sc~v})}{f(C{\sc~iv})}$ 
& R & $\Delta \rm R$ & n & T & U
& $\epsilon$     \\

\hline 

1 & $-1060$ & $114$ & $9.8$ & $104.2$ & $1.39$ & $18.8$ & $18.1$  & $5.0$ 
& $4.87$ & $-0.03$ & $-1.8$\\

2 &  $-655$ &  $45$ & $5.4$ &  $39.5$ & $0.61$ & $19.2$ & $18.3$  & $5.4$ 
& $4.37$ & $-1.29$ & $-5.6$\\

3 &  $-518$ &  $90$ & $9.6$ &  $80.4$ & $1.37$ & $19.4$ & $18.9$  & $3.8$ 
& $4.86$ & $-0.05$ & $-1.4$\\

4 &  $-344$ & $156$ & $5.4$ & $150.6$ & $0.63$ & $19.8$ & $19.8$  & $4.2$ 
& $4.36$ & $-1.23$ & $-4.8$\\

5 &  $-165$ &  $74$ & $5.9$ &  $68.7$ & $0.83$ & $20.4$ & $20.4$  & $2.6$ 
& $4.44$ & $-0.85$ & $-4.1$\\

\hline
\end{tabular}
\end{center}
\vspace{-0.2in}
 
$^{a}$Velocities are given in ${\rm km\ s^{-1}}$. All other entries 
are given as logarithm of the value expressed in cgs.

$^{b}$Values are from C99.

\newpage
\normalsize{
\centerline{\bf Table 4}
\centerline{\bf Component Ion Recombination Times}
\centerline{\bf and Percentage Total Column Contributions}
\centerline{\bf for NGC~5548}
\medskip
\begin{center}
\begin{tabular}{ccccccc}
\hline\hline
\multicolumn{1}{c}{ n } 
& \multicolumn{1}{c}{C{\sc~iv}$^{a}$} 
& \multicolumn{1}{c}{\%$^{b}$}
& \multicolumn{1}{c}{O{\sc~vii}}
& \multicolumn{1}{c}{\%}
& \multicolumn{1}{c}{O{\sc~viii}} 
& \multicolumn{1}{c}{\%} \\

\hline 

1 & $2.71$ &  $7.59$  &  $5.39$  & $45.71$ & $6.62$ & $47.86$  \\

2 & $4.23$ &  $6.76$  &  $5.46$  &  $0.03$ & $5.54$ & $>0.01$ \\

3 & $3.96$ & $10.00$  &  $6.60$  & $53.70$ & $7.80$ & $52.48$  \\

4 & $5.30$ & $69.18$  &  $6.59$  &  $0.38$ & $6.57$ &  $0.01$  \\

5 & $6.39$ &  $7.41$  &  $8.14$  &  $0.31$ & $8.30$ &  $0.02$  \\

\hline

$\Sigma ^{c}$ & $5.51$   &        &  $6.43$ &      &  $7.54$ &    \\

\hline

\end{tabular}
\end{center}
\medskip

$^{a}$ Values are the $\log t(s)$ of the absolute value of the
mean component recombination time. All C{\sc~iv} responses are 
anticorrelated and all O{\sc~vii} and O{\sc~viii} 
responses are correlated with a continuum drop.

$^{b}$ Percentage of the total ionic column density.

$^{c}$ Values are the $\log t(s)$ of the absolute value of the
mean recombination time averaged over all components and weighted by their 
relative column density contributions.
 
\end{document}